\documentclass[pra,twocolumn,floatfix]{revtex4}
\usepackage{graphicx}
\usepackage{color}
\usepackage{amsmath,amssymb,mathtools}
\usepackage{lipsum}
\DeclareUnicodeCharacter{2009}{\,}
\DeclarePairedDelimiter\abs{\lvert}{\rvert}

\begin{document}

\title{Heterosymmetric states of rotating quantum droplets under confinement}

\author{S. Nikolaou$^1$, G. M. Kavoulakis$^{1,2}$, and M. \"{O}gren$^{2,3}$}
\affiliation{$^1$Department of Mechanical Engineering, Hellenic Mediterranean University, P.O. Box 1939, 71004, Heraklion, Greece
\\
$^2$HMU Research Center, Institute of Emerging Technologies, 71004, Heraklion, 
Greece
\\
$^3$School of Science and Technology, \"{O}rebro University, 70182, \"{O}rebro, Sweden}
\date{\today}

\begin{abstract}

We investigate the rotational response of a confined, two-dimensional quantum droplet, which emerges in an attractive binary Bose mixture that is stabilized against collapse by beyond-mean-field effects. We consider both a harmonic and an anharmonic form for the external confining potential. We go beyond the widely employed ``phase-locked" single-order-parameter model, maintaining two separate order parameters for the two components, and calculating the lowest-energy state for various values of the angular momentum. For a population-balanced quantum droplet and sufficiently tight confinement, we find that near certain half-integer values of the angular momentum the droplet is excited in a ``heterosymmetric" manner, with the two components carrying different vorticities. This mode is naturally missed by the single-order-parameter model. We additionally investigate the effects of a small population imbalance in the droplet. Apart from an energy increase associated with the population difference, the imbalance also lifts the double degeneracy of the heterosymmetric states, which characterizes the $\mathbb{Z}_2$-symmetric balanced droplet. The heterosymmetric mode is found to be favored by the energy term which captures the beyond-mean-field effects in the mixture.

\end{abstract}

\maketitle

\section{Introduction}
Ever since the first dilute atomic Bose-Einstein condensates were realized experimentally roughly 30 years ago \cite{Cornell, Ketterle, Hulet}, the research in the field of cold atomic gases has been expanding in various different directions. A subfield that has attracted significant attention is that of mixtures of distinguishable atoms \cite{Mixture1, Mixture2, Mixture3}. Quite generally, one can argue that the additional degrees of freedom arising from the different components, even in the simplest, two-component case, give rise to a variety of effects which are absent in single-component gases \cite{Mixture4, Mixture5}.

An interesting example of how the binary character of a mixture gives rise to novel physics has been demonstrated theoretically by Petrov \cite{Petrov}. In his seminal work, it was shown that a two-component mixture with attractive intercomponent interactions in the collapse threshold, unstable from the mean-field point of view, is stabilized by the effectively repulsive Lee-Huang-Yang energy term \cite{LHY}. This term describes beyond-mean-field effects, namely, the fluctuations of the Bogoliubov vacuum, which are normally negligible. Via this mechanism, a self-bound object arises, called a quantum droplet. The description of this mechanism was subsequently extended to lower dimensions \cite{PetrovAstra}, and the emergence of quantum droplets was also demonstrated experimentally \cite{DropExp1, DropExp2, DropExp3, DropExp4, DropExp5, DropExp6}. This novel system has attracted a lot of attention in the literature. Here, we point to the review articles \cite{DropRev1,DropRev2,DropRev3}.

The rotational response of quantum droplets has been a subject of extensive study, naturally driven by their superfluid character. One of the characteristic properties of superfluidity is the formation of quantized vortices when the superfluid is set to rotation \cite{Leggett}. Quantum droplets indeed exhibit the superfluid sign of quantized vortices \cite{DropVor1,DropVor2,DropVor3,DropVor4,DropVor5,EK,Kartashov,DropVor6,Caldara,DropVor7,NKO1,DropVor8,NKO2,DropVor9,NKO3,DropVor10,Poparic}, and they are even found to manifest a mixed rotational state, combining vortex and center-of-mass excitation \cite{NKO1,NKO3}, which does not appear in single-component condensates.

It is interesting to note that the rotational properties of quantum droplets, a mixture with attraction between its two components, have a striking difference to the corresponding case of intercomponent repulsion. In the latter case, the rotating states generally tend to have one component localized in the vortex core(s) of the other component, e.g., by forming coreless vortices \cite{Coreless1,Coreless2,Coreless3,Coreless4,Coreless5,Coreless6,Coreless7,Coreless8}. This arises as a consequence of the minimization of the interaction energy, which favors phase separation in the densities of the two components \cite{Mixture5,PhaseSep1,PhaseSep2}. Conversely, in the former case, the minimization of the interaction energy favors the overlap of the two components \cite{Mixture5}, with the mixture tending to behave similarly to a single-component condensate, especially in the case of equal masses. This tendency is exploited widely in the study of quantum droplets, where the system of the coupled, nonlinear differential equations that describe the system can be reduced to a single equation, describing a single order parameter which is shared between the two components \cite{Petrov,PetrovAstra}. The majority of works on rotating quantum droplets utilize this reduced model. Therefore, the two components of a rotating droplet have usually been found to be excited in an identical manner.

The above is not to suggest that the physics of mixtures with intercomponent attraction can always be reduced to the picture of a single-component gas. For example, a rotating heteronuclear two-component mixture, with attractive intercomponent contact interactions, has been found to exhibit exotic vortex lattice structures in the ground state, with the components manifesting different lattice geometries \cite{Mottonen}. In the case of quantum droplets, a heteronuclear system has been found to host vortices in only the most massive component \cite{Caldara}, with the lighter component hosting partially filled cores that do not carry actual vorticity. Interestingly, such structures have also been predicted in homonuclear quantum droplets with an intercomponent interaction asymmetry \cite{Poparic}.

Therefore, the question that arises, and the one we aim to answer in this work, is with regard to cases where the reduced, single-order-parameter model of a rotating droplet is insufficient to correctly describe the system. In other words, we seek to determine under what circumstances the two components of the rotating droplet are excited in a different manner. We place an emphasis on ``heterosymmetric" vortex states, where the two components carry different vorticities. Such states have been investigated in Ref.\,\cite{Kartashov}, for the case of unconfined droplets.

In this work, we focus on two-dimensional, confined quantum droplets, with equal masses in the two components and equal intercomponent interaction strengths. We consider both cases of harmonic and anharmonic confinement. We demonstrate that, for certain values of the angular momentum and sufficiently tight confinement, the lowest-energy states are heterosymmetric vortex states, which are naturally missed by the single-order-parameter model. We determine that the emergence of heterosymmetric lowest-energy states is due to the competition among the various energy terms in the system, with such configurations being favored by the beyond-mean-field energy contribution. We also consider the effects of a small population imbalance in the droplet. Interestingly, we find heterosymmetric lowest-energy states even in the absence of a population imbalance, thus their appearance cannot be attributed solely to a mass, interaction or population asymmetry.

The remainder of the paper is organized as follows. In Sec.\,II we introduce the model we use, where we maintain two separate order parameters for the two components. In Sec.\,III we present our numerical results for the case of harmonic confinement, both for a balanced and an imbalanced quantum droplet. We also present a semianalytic calculation that supports these results. In Sec.\,IV we present our numerical results for the case of anharmonic confinement, again for a balanced and an imbalanced quantum droplet. Finally, in Sec.\,V we summarize and discuss the main results of this work.

\section{Model}
We assume a binary ($i=\;\uparrow,\downarrow$) mixture with equal masses of the atoms in the two components, $m_\uparrow=m_\downarrow=m$. We also assume that the mixture is confined to two dimensions via a very tight harmonic potential along the axis of rotation. The atoms are interacting weakly via short-range potentials, which are determined by the coupling constants $g_{\uparrow\uparrow}$ and $g_{\downarrow\downarrow}$ for the repulsive intraspecies interactions, and the coupling constant $g_{\uparrow\downarrow}$ for the attractive interspecies interaction. Setting $m = \hbar = 1$, the energy density functional is \cite{PetrovAstra}
\begin{multline}
\mathcal{E}[\Psi_\uparrow,{\Psi_\uparrow}^{\negthickspace*},\Psi_\downarrow,{\Psi_\downarrow}^{\negthickspace*}] = \frac{1}{2} \left( \sqrt{g_{\uparrow\uparrow}}\abs{\Psi_\uparrow}^2 - \sqrt{g_{\downarrow\downarrow}}\abs{\Psi_\downarrow}^2 \right)^2 \\ + \frac{1}{8\pi} \left( g_{\uparrow\uparrow}\abs{\Psi_\uparrow}^2 + g_{\downarrow\downarrow}\abs{\Psi_\downarrow}^2 \right)^2 \\ \times \ln \frac{\left( g_{\uparrow\uparrow}\abs{\Psi_\uparrow}^2 + g_{\downarrow\downarrow}\abs{\Psi_\downarrow}^2 \right)\sqrt{e}}{\Delta} \,.
\label{energy_density_general}
\end{multline}
Here, $\Delta$ is an energy parameter introduced so that the (regularized) energy density does not explicitly depend on the momentum cutoff. Its value is
\begin{equation}
\Delta = \sqrt{\epsilon_{\uparrow\downarrow}\sqrt{\epsilon_{\uparrow\uparrow}\epsilon_{\downarrow\downarrow}}} \exp \biggl[ \frac{-\ln^2 ( \epsilon_{\uparrow\uparrow}/\epsilon_{\downarrow\downarrow} )}{4\ln ( \epsilon_{\uparrow\uparrow}\epsilon_{\downarrow\downarrow}/\epsilon_{\uparrow\downarrow}^2 ) } \biggr] \,,
\label{Delta_general}
\end{equation}
chosen such that $\delta g = g_{\uparrow\downarrow} + \sqrt{g_{\uparrow\uparrow}g_{\downarrow\downarrow}} = 0$, i.e., the mixture is on the collapse threshold. Here, $\epsilon_{ii'} = 4e^{-2\gamma}/a_{ii'}^2$, where $\gamma$ is the Euler-Mascheroni constant and $a_{ii'}$ are the two-dimensional scattering lengths, related to the three-dimensional ones via the formula
\begin{equation}
\epsilon_{ii'} = \frac{b}{\pi a_z^2}\exp \Bigl( \sqrt{2\pi}a_z/a_{ii'}^{\rm 3D} \Bigr) \,,
\end{equation}
where $a_z$ is the oscillator length in the confinement direction and $b \approx 0.915$ \cite{PetrovShlyapnikov}. Substituting $\epsilon_{ii'}$, one gets
\begin{equation}
a_{ii'}^2 = \frac{4\pi a_z^2 e^{-2\gamma}}{b} \exp \Bigl( -\sqrt{2\pi}a_z/a_{ii'}^{\rm 3D} \Bigr) \,.
\label{2Dlengths}
\end{equation}
Additionally, the coupling constants for the intra- and intercomponent interactions are given as
\begin{equation}
g_{ii'} = \frac{4\pi}{\ln ( \epsilon_{ii'}/\Delta )} \,.
\label{coupling_constant_general}
\end{equation}

Values of the two component densities $n_i = \abs{\Psi_i}^2$ that give rise to a self-bound quantum droplet, that is, equilibrium with the vacuum, lie close to the line $n_\uparrow/n_\downarrow = \sqrt{g_{\downarrow\downarrow}/g_{\uparrow\uparrow}}$ \cite{Petrov,PetrovAstra}. In the balanced case, the two component densities are locked to this ratio, for which the first term in Eq.\,(\ref{energy_density_general}) vanishes. If one additionally neglects any out-of-phase motion between the components, the system can be described in terms of a single order parameter, $\Psi = \sqrt{2}\Psi_\uparrow = \sqrt{2}\Psi_\downarrow$. Thus, the single-order-parameter model neglects fluctuations of the spin channel, as well as vortex configurations characterized by different vorticities in the two components. Here, we will go beyond the single-order-parameter model, maintaining two separate order parameters $\Psi_i$.

We assume an interaction-symmetric system, $g_{\uparrow\uparrow} = g_{\downarrow\downarrow} = g$. Equivalently, $a_{\uparrow\uparrow} = a_{\downarrow\downarrow} = a$ and $\epsilon_{\uparrow\uparrow} = \epsilon_{\downarrow\downarrow} = \epsilon$. Here, the balanced case reduces to $n_\uparrow = n_\downarrow$, i.e., any imbalance in the droplet will arise due to an imposed population difference. Note that an interaction-symmetric balanced droplet is associated with a $\mathbb{Z}_2$ symmetry, being invariant under the exchange of the $\uparrow$ and $\downarrow$ atoms \cite{Z2symmetry}. Under this assumption, we have
\begin{equation}
\Delta = \sqrt{\epsilon_{\uparrow\downarrow}\epsilon} = \frac{4e^{-2\gamma}}{a_{\uparrow\downarrow}a}
\label{Delta_symmetric}
\end{equation}
and
\begin{equation}
g = \frac{4\pi}{\ln ( a_{\uparrow\downarrow}/a )} \,.
\label{coupling_constant_symmetric}
\end{equation}
It follows from Eq.\,(\ref{2Dlengths}) that
\begin{equation}
\ln (a_{\uparrow\downarrow}/a) = \sqrt{\frac {\pi} 2} 
\biggl( \frac {a_z} {a^{\rm 3D}} - \frac {a_z} {a_{\uparrow\downarrow}^{\rm 3D}} \biggr) \,.
\label{logarithm}
\end{equation}
We also assume that the droplet is confined in a two-dimensional quadratic-plus-quartic potential
\begin{equation}
V(\rho) = \frac 1 2 \omega^2 \rho^2 \left( 1 + \lambda \frac {\rho^2} {a_0^2} \right) \,,
\label{trapping_potential}
\end{equation} 
where $\rho$ is the radial coordinate, $\omega$ is the frequency of the harmonic potential, $a_0 = \sqrt{1/\omega}$ is the oscillator length, and $\lambda$ is a dimensionless parameter which controls the strength of the anharmonic part of the trapping potential. When $\lambda = 0$ this potential reduces to the familiar harmonic form, $V(\rho) = \frac 1 2 \omega^2 \rho^2$, which we use in a large part of this work.

Let us introduce the following units. Here, we reinsert $m$ and $\hbar$, in order to make the definition of the units more transparent. First, the unit of density is
\begin{equation}
\Psi_0^2 = \frac {e^{-2 \gamma - 1}\ln ( a_{\uparrow\downarrow}/a )} {\pi a_{\uparrow\downarrow}a} \,.
\end{equation}
The unit of length is
\begin{equation}
x_0 = \sqrt{\frac {a a_{\uparrow\downarrow} \ln(a_{\uparrow\downarrow}/a)} 
{4 e^{-2 \gamma - 1}}} \,,
\label{length_unit}
\end{equation}
while those of the energy $E_0$ and of the frequency $\omega_0$ are  
\begin{equation}
E_0 = \hbar \omega_0 = \frac {\hbar^2} {m x_0^2} = \frac {\hbar^2 4 e^{-2 \gamma - 1}} {m a a_{\uparrow\downarrow} \ln(a_{\uparrow\downarrow}/a)} \,.
\end{equation}
Finally, the number of atoms is measured in units of $N_0$, where
\begin{equation}
N_0 = \Psi_0^2 x_0^2 = \frac {\ln^2(a_{\uparrow\downarrow}/a)} {4 \pi} \,.
\label{atom_number_unit}
\end{equation}
In the rest of the paper we work in dimensionless units, using the units presented above.

The extended energy functional takes the form
\begin{multline}
E[\Psi_\uparrow,{\Psi_\uparrow}^{\negthickspace*},\Psi_\downarrow,{\Psi_\downarrow}^{\negthickspace*}] = \int \biggl[ \frac{1}{2}\abs{\nabla\Psi_\uparrow}^2 + \frac{1}{2}\abs{\nabla\Psi_\downarrow}^2 \\
+ V\left( \abs{\Psi_\uparrow}^2 + \abs{\Psi_\downarrow}^2 \right) + \frac{D}{2} \left( \abs{\Psi_\uparrow}^2 - \abs{\Psi_\downarrow}^2 \right)^2 \\
+ \frac{1}{2} \left( \abs{\Psi_\uparrow}^2 + \abs{\Psi_\downarrow}^2 \right)^2 \ln \frac{ \abs{\Psi_\uparrow}^2 + \abs{\Psi_\downarrow}^2 }{\sqrt{e}} \biggr] \,d^2 r \\
- \mu_\uparrow \int {\Psi_\uparrow}^{\negthickspace*} \Psi_\uparrow \,d^2 r - \mu_\downarrow \int {\Psi_\downarrow}^{\negthickspace*} \Psi_\downarrow \,d^2 r \\
- \Omega \int \bigl( {\Psi_\uparrow}^{\negthickspace*} {\hat L} \Psi_\uparrow + {\Psi_\downarrow}^{\negthickspace*} {\hat L} \Psi_\downarrow \bigr) \,d^2 r \,,
\label{energy_functional}
\end{multline}
where we have set $D = \ln(a_{\uparrow\downarrow}/a)$, as defined by Eq.\,(\ref{logarithm}). Additionally, $\mu_i$ are the chemical potentials of the two components, $\Omega$ is the rotational frequency and ${\hat L}$ is the angular momentum operator. Here, $\Psi_i$ are normalized to the (scaled) number of atoms in each component, $\int \abs{\Psi_i}^2 \,d^2 r = N_i$. We stress that throughout this work, when we refer to the number of atoms we will mean the scaled number, using the unit of Eq.\,(\ref{atom_number_unit}), unless noted otherwise. In the following, we will write $N = N_\uparrow + N_\downarrow$ for the total number of atoms, and $\delta N = N_\uparrow - N_\downarrow > 0$ for the population imbalance (thus, $\uparrow$ and $\downarrow$ will refer to the majority and minority component, respectively). In addition, we will refer to the first nonlinear energy term as the ``contact" term, and the second one as the ``logarithmic" term. The corresponding coupled nonlinear Schr\"odinger equations that are satisfied by $\Psi_i$ are \cite{DropVor2,NLSEs1,NLSEs2,NLSEs3}
\begin{multline}
\mu_i \Psi_i = \biggl[ -\frac{1}{2} \nabla^2 + V + D \left( \abs{\Psi_i}^2 - \abs{\Psi_{i'}}^2 \right) \\
+ \left( \abs{\Psi_i}^2 + \abs{\Psi_{i'}}^2 \right) \ln  \left( \abs{\Psi_i}^2 + \abs{\Psi_{i'}}^2 \right) - \Omega {\hat L} \biggl] \Psi_i \,.
\label{NLSEs}
\end{multline}
The coupled equations are solved by minimizing numerically the functional of Eq.\,(\ref{energy_functional}), using the damped second-order-in-fictitious-time method \cite{DFPM}, which is a method for constrained minimization. We work under the constraints of fixed number of atoms $N_i$ in each component and fixed total angular momentum $L$. In this scheme, $\mu_i$ and $\Omega$ are Lagrange multipliers, corresponding to the conservation of the atom number in each component and of the total angular momentum, respectively. We use a square spatial grid, with $\delta x = \delta y = 0.125$, which proves to be accurate enough, in the sense that it produces results that are converged with respect to the grid resolution. Also, the size of the calculation domain is larger than presented in the figures below, in order to avoid boundary effects.

Finally, we stress that we use a variety of trial order parameters as initial conditions for the calculations, considering trial functions that are the same for the two components, as well as different between them. The use of multiple initial conditions in the calculation, for each value of the angular momentum, and the comparison of the corresponding energies of the solutions, are necessary to verify that we have reached the lowest-energy state and not some local minimum of the energy functional, which would correspond to an excited, metastable state. 

\section{Yrast states of harmonically confined quantum droplets}
We initially focus on rotating quantum droplets confined in a harmonic potential, setting $\lambda = 0$ in Eq.\,(\ref{trapping_potential}). We generally set $\omega = 0.05$ for the trapping frequency, in order to facilitate a comparison with the results we have obtained using the single-order-parameter model in Ref.\,\cite{NKO1}. Apart from being a common form for the external trapping, a harmonic potential also allows for a certain degree of analytic treatment, which is useful in supporting the numerical calculations. We consider a range of angular momentum values and focus on the yrast states, that is, the lowest-energy state for each value of $L$. Additionally, we consider both balanced and imbalanced quantum droplets. It is instructive to first give a brief overview of some results regarding the low-angular-momentum limit $\ell \ll 1$, where $\ell = L/N$, before moving to vortex states.

\subsection{Center-of-mass and surface-wave excitation}
A possible rotating state for a harmonically confined quantum droplet is center-of-mass motion with respect to the trap axis, while maintaining its internal correlations unaffected. This excitation mode involves a dispersion relation, that is, the energy as a function of the angular momentum $E(L)$, which is linear, with a slope equal to the trapping frequency. This is a well known result for the harmonic potential, in which the center-of-mass coordinate separates from the relative coordinates. In fact, for sufficiently weak confinement, this rotating state appears as the yrast state, whereas beyond this limit the angular momentum is carried by surface waves in the lowest-energy state \cite{NKO1}.

\begin{figure}
\centering
\includegraphics[width=\columnwidth]{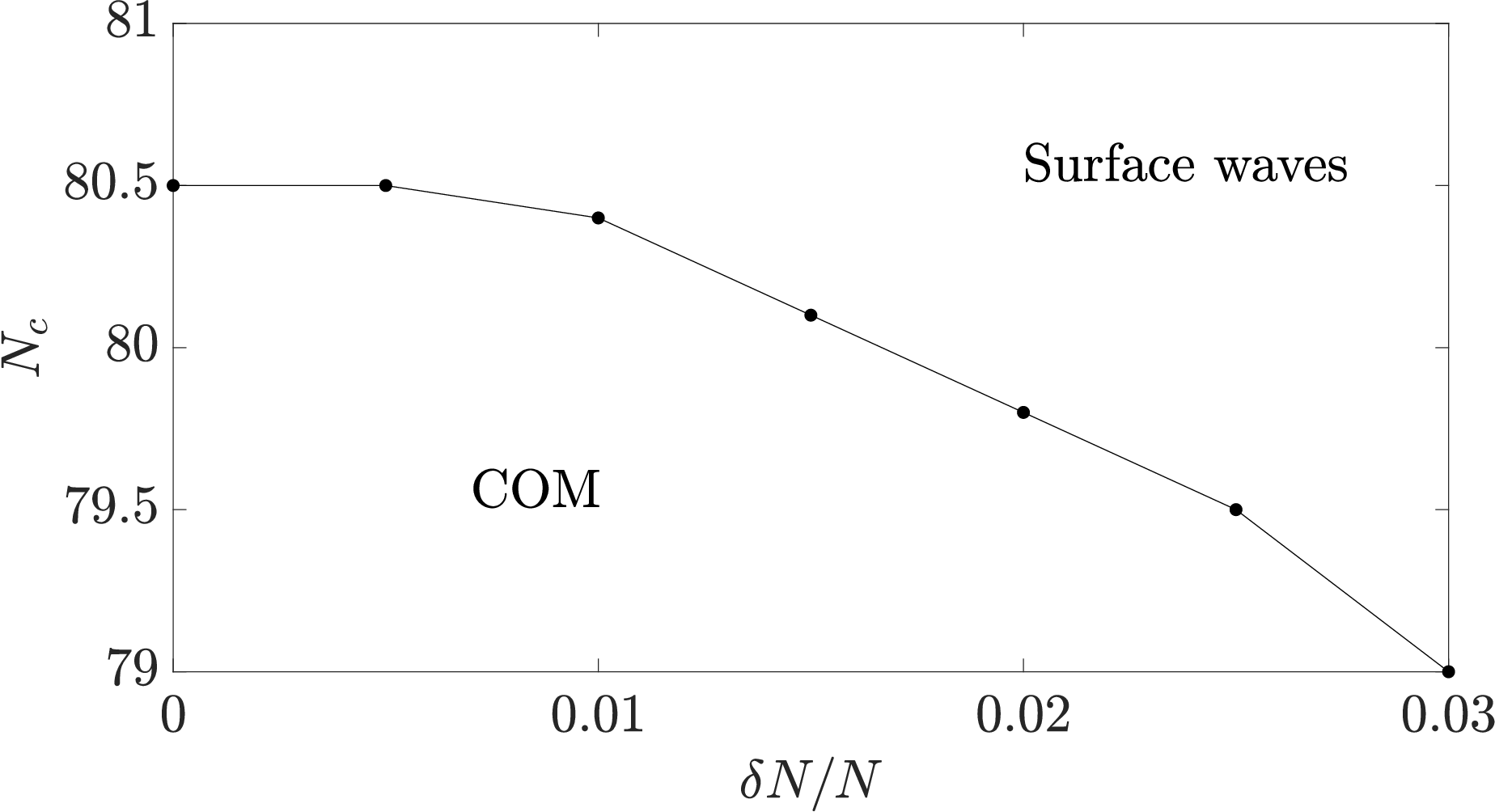}
\caption{The critical value of the total number of atoms $N$ for which the droplet carries its angular momentum via surface waves instead of center-of-mass motion, as a function of the population imbalance ratio $\delta N/N$. Here $\omega=0.05$ and $D=25$. This plot may also be viewed as the low-angular-momentum phase diagram (produced for $\ell=0.2$) for an imbalanced droplet, involving center-of-mass excitation below the curve, and surface-wave excitation above it.}
\end{figure}

Let us recall here that there are two relevant length scales in the problem. One is defined by the oscillator length of the harmonic potential, $a_0 = \sqrt{1/\omega}$, and the other by the droplet radius, $\rho_0 \propto \sqrt{N}$ \cite{EK,NKO1}. Roughly speaking, a weakly confined droplet satisfies $\rho_0 \ll a_0$, that is, $N$ or $\omega$ is sufficiently small, whereas a strongly confined droplet satisfies $\rho_0 \gg a_0$, that is, $N$ or $\omega$ is sufficiently large. Equivalently, we may say that the value of $N\omega$ is a measure of the confinement strength for the droplet, with larger values corresponding to stronger confinement. In particular, in Ref.\,\cite{NKO1} we have found, for a balanced droplet within the single-order-parameter model, that the critical value of $N\omega$ below which the droplet performs center-of-mass rotation is $\approx 4$. More specifically, for $\omega=0.05$, the corresponding critical value of the total number of atoms, which marks the transition to surface-wave excitation, is $N_c = 80.5$ \footnote{In Refs.\,\cite{NKO1,NKO3} this value was erroneously reported to be $98.7$. The value presented in this work is the corrected value for the transition.}. Here, we found that these results are reproduced exactly by the present model, which includes two order parameters. This agreement also serves as a check for the numerical implementation of the model.

Regarding the case of surface-wave excitation, we have found that the value of $N\omega$ also plays a role in determining the excitation order $k$, in terms of a $2^k$-pole excitation, that is, the shape of the deformation of the rotating droplet \cite{KMP}. Note that $k$ corresponds to the number of vortices surrounding the rotating droplet. In particular, as $N\omega$ increases, the yrast state turns from center-of-mass motion (which, in this scheme, is associated with the dipole mode, $k=1$), to surface waves with a quadrupole deformation, $k=2$. Then, as $N\omega$ increases further, the surface-wave excitation involves an octupole deformation, $k=3$, then a hexadecapole ($16$-pole) deformation, $k=4$, etc.

Within the present model, we also investigated the effects of a nonzero population imbalance in the low-angular-momentum limit. We found that the physical picture is qualitatively the same as for a balanced quantum droplet, apart from a density difference between the two components. Regarding the crossover from center-of-mass motion to surface waves, we found that the presence of imbalance slightly modifies the transition in a quantitative manner. Specifically, it induces a decrease of the critical value of $N$ for which the droplet turns to surface-wave excitation. This behavior is illustrated in Fig.\,1, which we have produced for $D = 25$ and $\omega=0.05$. We see that as the population imbalance increases, $N_c$ decreases, becoming $79$ for $\delta N/N = 0.03$. We stress that this decrease is small, corresponding to approximately $75$ atoms in physical units, as per Eq.\,(\ref{atom_number_unit}).

\subsection{Heterosymmetric vortex states of balanced droplets}
The presence of surface waves in the low-$L$ limit signifies the emergence of vortex states as $L$ increases. We have investigated such states in detail within the present model, particularly focused on configurations with different vorticity between the two components, that is, heterosymmetric states. For this reason, we are especially interested in values of $L$ that coincide with the population of either of the components. We initially focus on the case of a balanced droplet, with $N_\uparrow = N_\downarrow = N/2$.

As the angular momentum increases, $k-1$ phantom vortices in the periphery of the droplet recede while the remaining one approaches the droplet, progressively penetrating its surface. Within the single-order-parameter model, we find that the vortex approaches the center of the droplet as $L$ increases further, settling at the origin for $L = N$, in an axially symmetric configuration \cite{NKO1}. Throughout all these $L$ values, both components of the droplet are excited in the same manner. We will refer to this type of excitation as the ``phase-locked" mode. Within the present model, where we maintain two separate order parameters, a more exotic situation arises for values of $L \approx N/2$. As $L$ approaches this value, for certain parameter values, a heterosymmetric excitation mode becomes the yrast state, where a vortex appears only in one component, accompanied by a partially filled vortex core in the other component. In this case, the angular momentum is carried predominantly by the component carrying the vortex. We found numerically that this heterosymmetric state appears as an yrast state only for sufficiently high values of $N\omega$, i.e., for sufficiently strong confinement. In order to interpret these numerical findings, we developed a simple semianalytic model for the crossover from the phase-locked to the heterosymmetric state.

\subsubsection{A semianalytic model}
The energetic preference for the heterosymmetric states over the states with an off-center vortex in both components can be seen to arise mainly due to the interplay between the trapping potential and the nonlinear energy terms. In order to illustrate this, let us consider the single-particle eigenfunctions of the two-dimensional harmonic oscillator $\phi_{nm}(\rho,\theta)$, describing the Landau levels. Here, $n$ is the radial quantum number and $m$ is the quantum number corresponding to the angular momentum. In particular we utilize the states
\begin{equation}
\phi_{00} = \frac {1} {\sqrt{\pi} a_0} e^{-\rho^2/(2 a_0^2)} \,,
\end{equation}
\begin{equation}
\phi_{01} = \frac {1} {\sqrt{\pi} a_0^2} \rho e^{i \theta} e^{-\rho^2/(2 a_0^2)} \,,
\end{equation}
and
\begin{equation}
\phi_{10} = \frac {1} {\sqrt{\pi} a_0} \biggl(1-\frac{\rho^2}{a_0^2}\biggr) e^{-\rho^2/(2 a_0^2)} \,.
\end{equation}
We write $N_\uparrow = (N + \delta N)/2$ and $N_\downarrow = (N - \delta N)/2$. Here, we will treat the general case of nonzero imbalance, before setting $\delta N = 0$ to obtain the final result. We develop the model for states with angular momentum $L = N_\downarrow$, or equivalently, angular momentum per particle $\ell = L/N = 1/2 - \delta N/(2N)$. For the phase-locked state, where a vortex penetrates the surface of both components, we write
\begin{equation}
\Psi_{i,\mathrm{pl}} = \sqrt{N_i}\Bigl(\sqrt{1-\ell}\phi_{00}+\sqrt{\ell}\phi_{01}\Bigr)\,,
\end{equation}
where $i=\;\uparrow,\downarrow$. For the heterosymmetric state, with a (central) vortex only in the minority component, we write
\begin{equation}
\Psi_{\uparrow,\mathrm{h}} = \sqrt{N_\uparrow}\Bigl(\sqrt{d_0}\phi_{00} - \sqrt{d_1}\phi_{10}\Bigr)
\end{equation}
and
\begin{equation}
\Psi_{\downarrow,\mathrm{h}} = \sqrt{N_\downarrow}\phi_{01}\,.
\end{equation}
In the case of the phase-locked state, the values of the coefficients are fixed by the normalization and the expectation value of the angular momentum we have chosen. For the $\uparrow$ component of the heterosymmetric state the normalization condition is $d_0 + d_1 = 1$. Here we cannot employ a constraint based on the angular momentum value, as the states $\phi_{00}$ and $\phi_{10}$ are both independent of $\theta$, that is, the angular momentum of $\Psi_{\uparrow,\mathrm{h}}$ is zero, regardless of the values of $d_0$ and $d_1$. Instead, we treat the second constraint, which is necessary in order to determine both coefficients, in a variational manner. In particular, we choose them in such a way that the overlap between the component densities is maximized, that is, the value of $\int \bigl( \abs{\Psi_{\uparrow,\mathrm{h}}}^2 - \abs{\Psi_{\downarrow,\mathrm{h}}}^2 \bigr)^2 \,d^2 r$ is minimized. From a physical point of view, this constraint reflects the attraction between the two components, and corresponds to the minimization of the contact energy term, which has the smallest magnitude among the energy terms of the system. This constraint, coupled with the normalization condition, gives the values $d_0 = 0.74$ and $d_1 = 0.26$ for a balanced droplet.

Substituting $N_\downarrow$, $N_\uparrow$ and $\ell$, and employing the binomial approximation, which amounts to neglecting terms of order $(\delta N / N)^2$, we rewrite
\begin{equation}
\Psi_{\uparrow,\mathrm{pl}} = \frac{\sqrt{N}}{2} \biggl[ \biggl(1+\frac{\delta N}{N}\biggr)\phi_{00} + \phi_{01} \biggr]
\end{equation}
and
\begin{equation}
\Psi_{\downarrow,\mathrm{pl}} = \frac{\sqrt{N}}{2} \biggl[ \phi_{00} + \biggl(1-\frac{\delta N}{N}\biggr)\phi_{01} \biggr]
\end{equation}
for the phase-locked state, and
\begin{equation}
\Psi_{\uparrow,\mathrm{h}} = {\sqrt\frac{N}{2}} \biggl(1+\frac{\delta N}{2N}\biggr) \Bigl(\sqrt{d_0}\phi_{00} - \sqrt{d_1}\phi_{10}\Bigr)
\label{heterosymmetric_up_order_parameter}
\end{equation}
and
\begin{equation}
\Psi_{\downarrow,\mathrm{h}} = {\sqrt\frac{N}{2}} \biggl(1-\frac{\delta N}{2N}\biggr) \phi_{01}
\end{equation}
for the heterosymmetric state.

The total energy is 
\begin{multline}
E = \int \biggl[ \mathcal{E}_\mathrm{lin} + \frac{D}{2} \left( \abs{\Psi_\uparrow}^2 - \abs{\Psi_\downarrow}^2 \right)^2 \\
+ \frac{1}{2} \left( \abs{\Psi_\uparrow}^2 + \abs{\Psi_\downarrow}^2 \right)^2 \ln \frac{ \abs{\Psi_\uparrow}^2 + \abs{\Psi_\downarrow}^2 }{\sqrt{e}} \biggr] \,d^2 r \,,
\end{multline}
where $\mathcal{E}_\mathrm{lin}$ denotes the linear energy density terms, that is, the sum of the kinetic and potential energies
\begin{equation}
\mathcal{E}_\mathrm{lin} = \sum_{i=\uparrow,\downarrow} \biggl( \frac{1}{2}\abs{\nabla\Psi_i}^2 + V\abs{\Psi_i}^2 \biggr) \,.
\end{equation}
Substituting the order parameters, we obtain the energy of the phase-locked state
\begin{multline}
E_\mathrm{pl} = L\omega + \frac{N^2}{8 \pi^2 a_0^2} \biggl[ c_0^2 D I_{1,\mathrm{pl}} + \ln\biggl(\frac{N}{2\pi\sqrt{e}a_0^2}\biggr) I_{2,\mathrm{pl}} \\ - I_{3,\mathrm{pl}} + I_{4,\mathrm{pl}} \biggr]\,,
\label{phase-locked_energy}
\end{multline}
where
\begin{gather*}
I_{1,\mathrm{pl}} = \int_0^\infty \int_0^{2\pi} \rho e^{-2\rho^2} \left( 1 + \rho^2 + 2 \rho \cos\theta \right)^2 \,d\theta\,d\rho\,, \\
I_{2,\mathrm{pl}} = \int_0^\infty \int_0^{2\pi} \rho e^{-2\rho^2} \left( c_+ + c_-\rho^2 + 2 \rho \cos\theta \right)^2 \,d\theta\,d\rho\,, \\ 
I_{3,\mathrm{pl}} = \int_0^\infty \int_0^{2\pi} \rho^3 e^{-2\rho^2} \left( c_+ + c_-\rho^2 + 2 \rho \cos\theta \right)^2 \,d\theta\,d\rho\,,
\end{gather*}
and
\begin{multline}
I_{4,\mathrm{pl}} = \int_0^\infty \int_0^{2\pi} \rho e^{-2\rho^2} \left( c_+ + c_-\rho^2 + 2 \rho \cos\theta \right)^2 \\ \times \ln \left( c_+ + c_-\rho^2 + 2 \rho \cos\theta \right) \,d\theta\,d\rho\,.
\label{phase-locked_ints}
\end{multline}
Here, we have defined $c_0 = \delta N/N$, $c_+ = 1 + \delta N/N$ and $c_- = 1 - \delta N/N$. The $L\omega$ term in Eq.\,(\ref{phase-locked_energy}) is the contribution of the kinetic-plus-potential energy, relative to the ground-state energy of the harmonic potential. In addition, the term involving the integral $I_{1,\mathrm{pl}}$ represents the contact energy term, while the rest of the terms arise due to the logarithmic energy term. In the same fashion, we have for the heterosymmetric state
\begin{multline}
E_\mathrm{h} = L\omega + Nc_+d_1\omega + \frac{N^2}{8 \pi^2 a_0^2} \biggl[ D I_{1,\mathrm{h}} + \ln\biggl(\frac{N}{2\pi\sqrt{e}a_0^2}\biggr) I_{2,\mathrm{h}} \\ - I_{3,\mathrm{h}} + I_{4,\mathrm{h}} \biggr]\,,
\label{heterosymmetric_energy}
\end{multline}
where
\begin{gather*}
I_{1,\mathrm{h}} = \int_0^\infty \int_0^{2\pi} \rho e^{-2\rho^2} [F_-(\rho)]^2 \,d\theta\,d\rho\,, \\
I_{2,\mathrm{h}} = \int_0^\infty \int_0^{2\pi} \rho e^{-2\rho^2} [F_+(\rho)]^2 \,d\theta\,d\rho\,, \\ 
I_{3,\mathrm{h}} = \int_0^\infty \int_0^{2\pi} \rho^3 e^{-2\rho^2} [F_+(\rho)]^2 \,d\theta\,d\rho\,,
\end{gather*}
and
\begin{multline}
I_{4,\mathrm{h}} = \int_0^\infty \int_0^{2\pi} \rho e^{-2\rho^2} [F_+(\rho)]^2 \ln [F_+(\rho)] \,d\theta\,d\rho\,.
\label{heterosymmetric_ints}
\end{multline}
Here, for ease of notation, we have introduced the function
\begin{multline}
F_\pm(\rho) = c_+d_0 + c_+d_1\bigl(1-\rho^2\bigr)^2 - 2c_+\sqrt{d_0d_1}\bigl(1-\rho^2\bigr) \\ \pm c_-\rho^2 \,.
\end{multline}

We see that the contribution of the kinetic-plus-potential energy in the heterosymmetric state is increased by the term $Nc_+d_1\omega$, which arises due to the mixing of the second-to-lowest Landau level $\phi_{10}$. We stress that the use of the $\phi_{10}$ state in the linear combination of Eq.\,(\ref{heterosymmetric_up_order_parameter}) is crucial in describing the partially filled core in the $\uparrow$ component, that is, the density decrease close to $\rho = 0$. The state $\phi_{00}$ alone is not adequate to capture this behavior. In particular, if one wrote $\Psi_{\uparrow,\mathrm{h}} = \sqrt{N_\uparrow}\phi_{00}$ instead, the radial profile of $\Psi_{\uparrow,\mathrm{h}}$ would resemble a bright soliton residing inside the vortex of the $\downarrow$ component (in a situation similar to the coreless vortices that appear in binary mixtures with repulsive intercomponent interactions), whereas in reality, it resembles a so-called gray soliton \cite{GraySoliton1,GraySoliton2,GraySoliton3}. This reflects the fact that the two components do not phase-separate in the droplet system.

\begin{figure}
\centering
\includegraphics[width=\columnwidth]{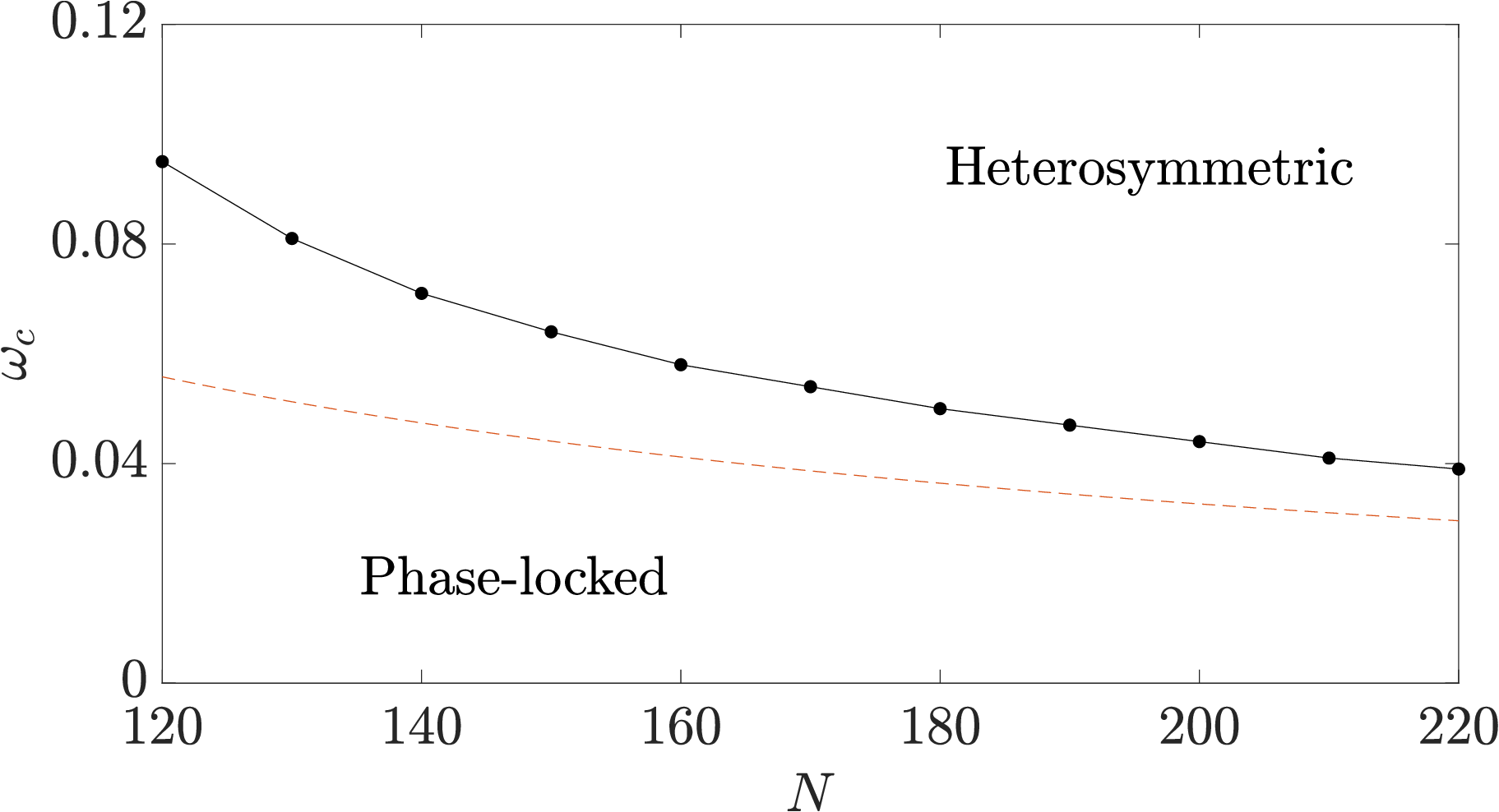}
\caption{Solid line, with data points: the critical value of the trapping frequency $\omega$ for which a heterosymmetric state, with a vortex only in the $\downarrow$ component, appears as the yrast state, as a function of the total number of atoms $N$. Here $L = N_\mathrm{\downarrow}$, $\delta N/N = 0$, and $D=25$. Dashed line: the semianalytic result we have derived for the crossover.}
\end{figure}

In order to investigate the crossover between the phase-locked and heterosymmetric states, let us set $E_\mathrm{pl} = E_\mathrm{h}$. We then have
\begin{multline}
\ln\biggl(\frac{N\omega}{2\pi\sqrt{e}}\biggr)(I_{2,\mathrm{pl}} - I_{2,\mathrm{h}}) = \frac{8\pi^2c_+d_1}{N} + D(I_{1,\mathrm{h}} - c_0^2 I_{1,\mathrm{pl}}) \\ - (I_{3,\mathrm{h}} - I_{3,\mathrm{pl}}) + I_{4,\mathrm{h}} - I_{4,\mathrm{pl}} \,,
\end{multline}
where we have substituted the oscillator length $a_0 = \sqrt{1/\omega}$. Finally, solving for $\omega$, we obtain the critical value of the trapping frequency for the crossover
\begin{equation}
\begin{split}
\omega_c =& \frac{2\pi}{N}\exp\biggl[\frac{1}{2} + \frac{8\pi^2c_+d_1}{N(I_{2,\mathrm{pl}}-I_{2,\mathrm{h}})} \\ &+ \frac{D(I_{1,\mathrm{h}}-c_0^2 I_{1,\mathrm{pl}})-(I_{3,\mathrm{h}}-I_{3,\mathrm{pl}})+I_{4,\mathrm{h}}-I_{4,\mathrm{pl}}}{I_{2,\mathrm{pl}}-I_{2,\mathrm{h}}} \biggr] \,.
\end{split}
\end{equation}

We have evaluated the integrals of Eqs.\,(\ref{phase-locked_ints}) and (\ref{heterosymmetric_ints}) numerically for $D = 25$ and $\delta N / N = 0$, that is, for the balanced case. In Fig.\,2 we present the critical value $\omega_c$ for the crossover, as a function of $N$, along with the fully numerical results. We see that our semianalytic model slightly underestimates the value of $\omega_c$. This is more pronounced for lower $N$, whereas the agreement improves as the number of atoms increases. This improvement is justified by the fact that the radius of the (unconfined) droplet increases with $N$, and thus, for any given $\omega$, the effects of the confinement also become more important with growing $N$, making the use of the Landau levels an increasingly better approximation. From an energetic point of view, in a weakly confined droplet, the logarithmic energy term is negative and the largest in magnitude. As $N$ increases, for a given $\omega$, the density of the droplet also increases and the logarithmic energy term moves towards zero, decreasing in magnitude, while the trapping potential energy term increases.

\begin{figure}
\centering
\includegraphics[width=\columnwidth]{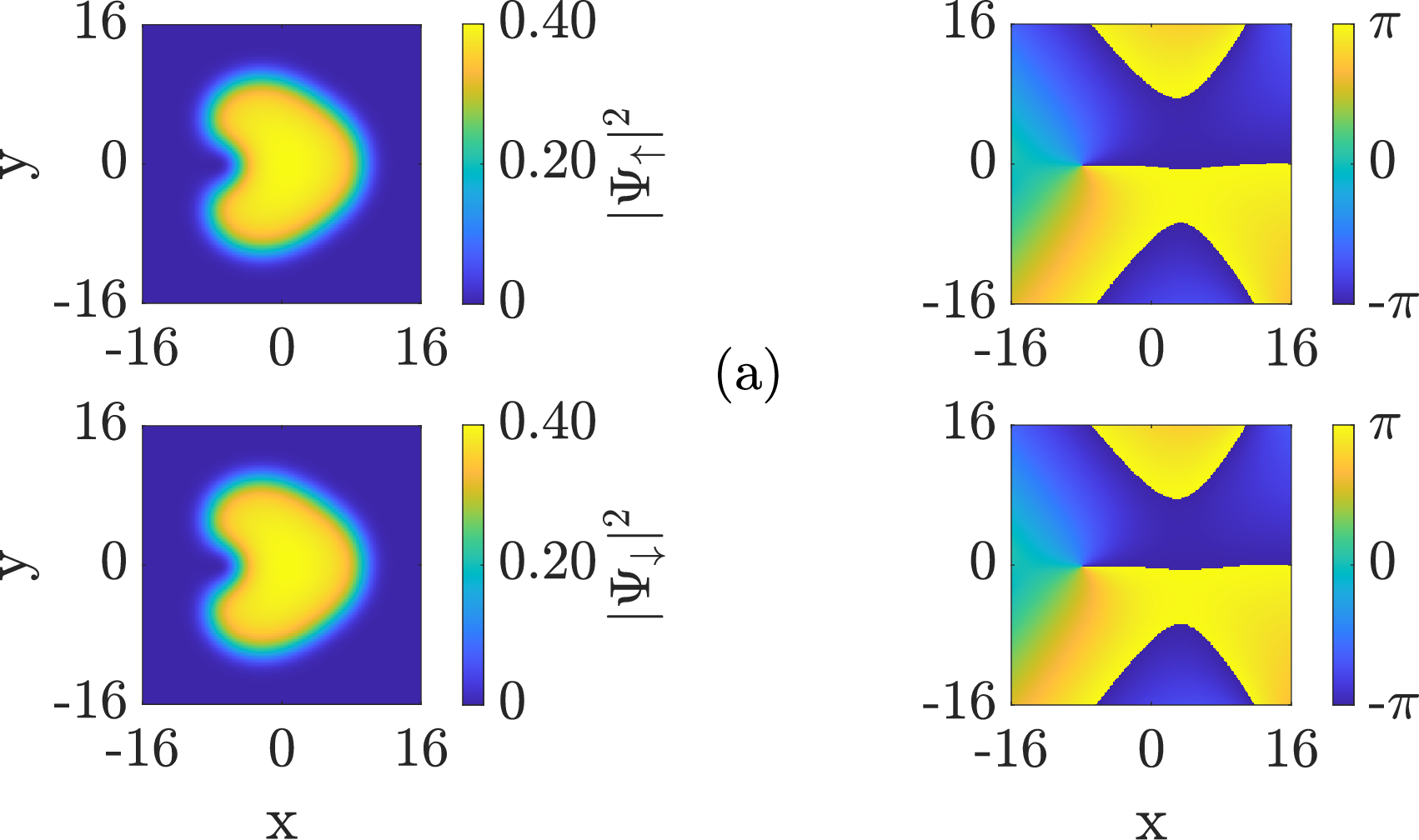}\\
\vspace{0.3\baselineskip}
\includegraphics[width=\columnwidth]{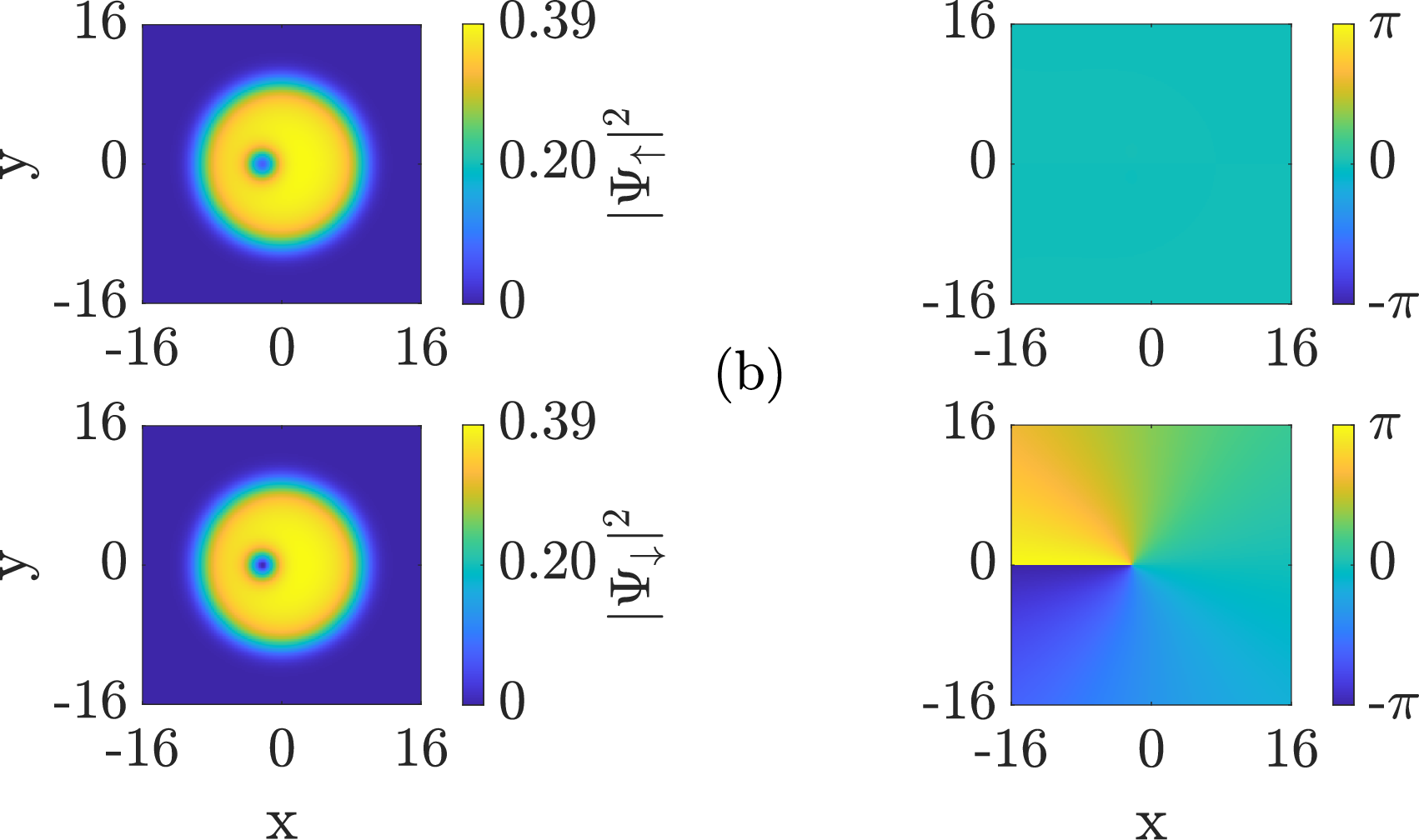}\\
\vspace{0.3\baselineskip}
\includegraphics[width=\columnwidth]{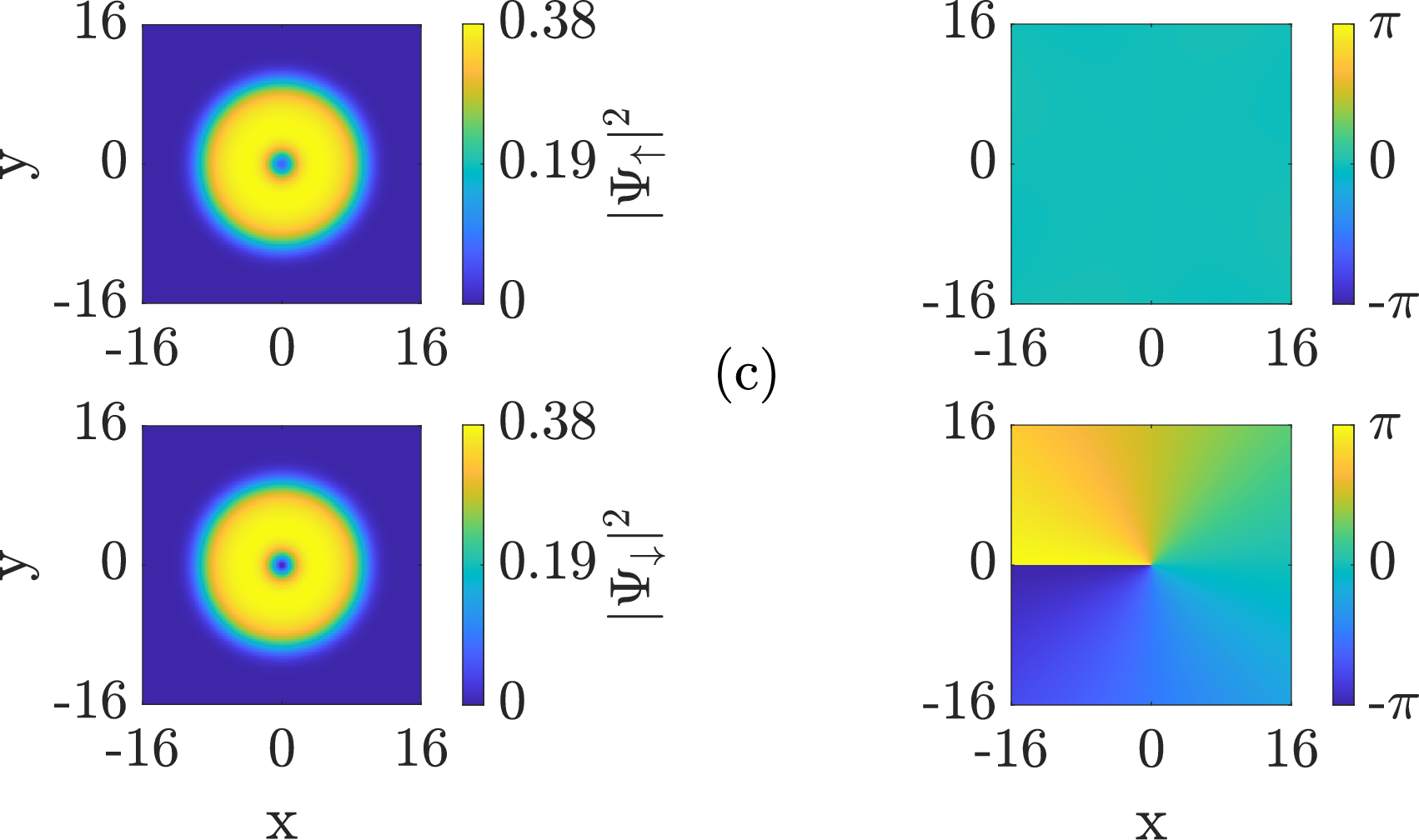}
\caption{(a)\textendash(c) The densities (left column, in units of $\Psi_0^2$) and the phases (right column) of the droplet order parameters, in the yrast state, for $N = 200$, $\delta N = 0$, $\omega = 0.05$, $D = 25$, and (a) $\ell = 0.4$, (b) $\ell = 0.47$, and (c) $\ell = 0.5$. The unit of length is $x_0$.}
\addtocounter{figure}{-1}
\end{figure}

\begin{figure}
\centering
\includegraphics[width=\columnwidth]{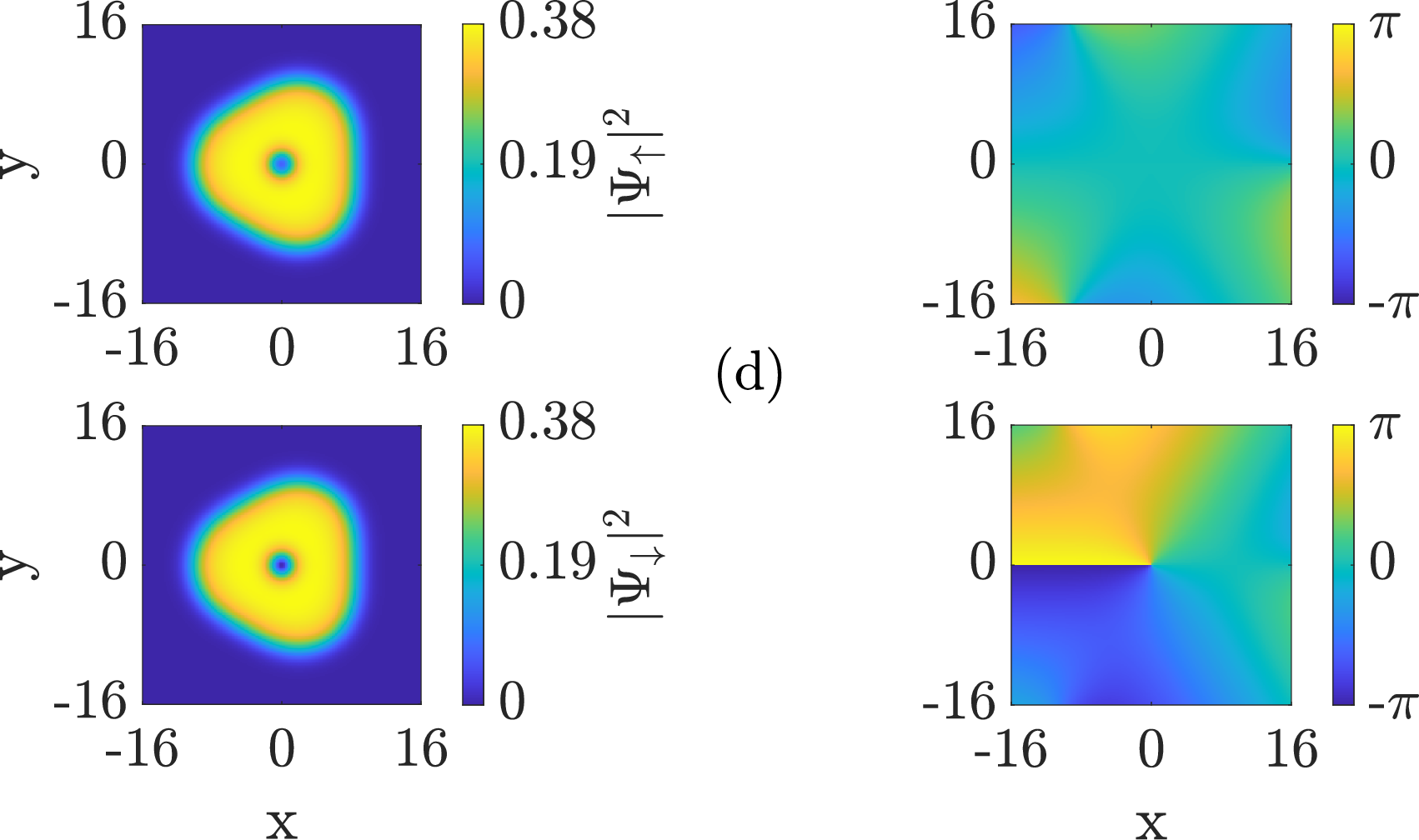}\\
\vspace{0.3\baselineskip}
\includegraphics[width=\columnwidth]{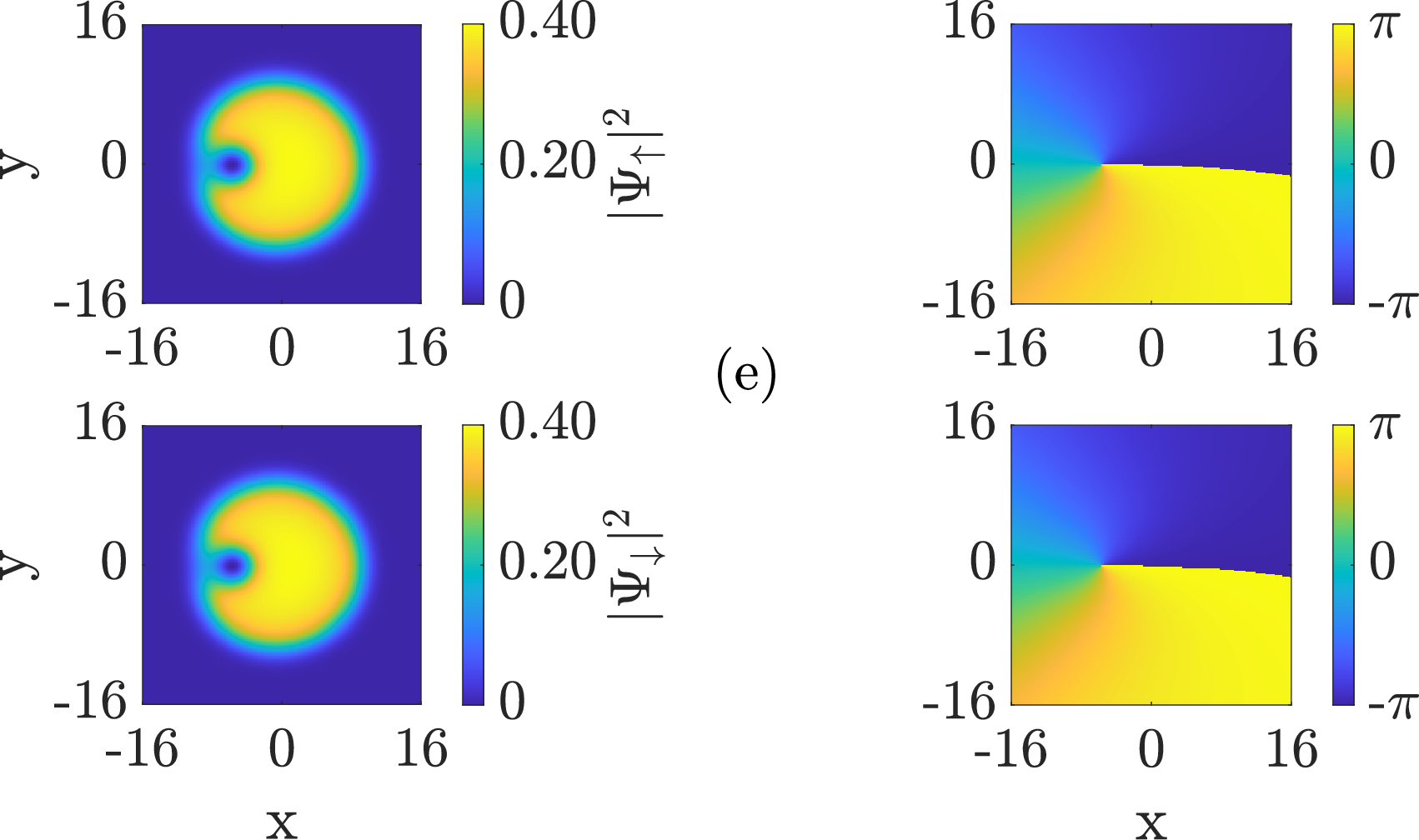}\\
\vspace{0.3\baselineskip}
\includegraphics[width=\columnwidth]{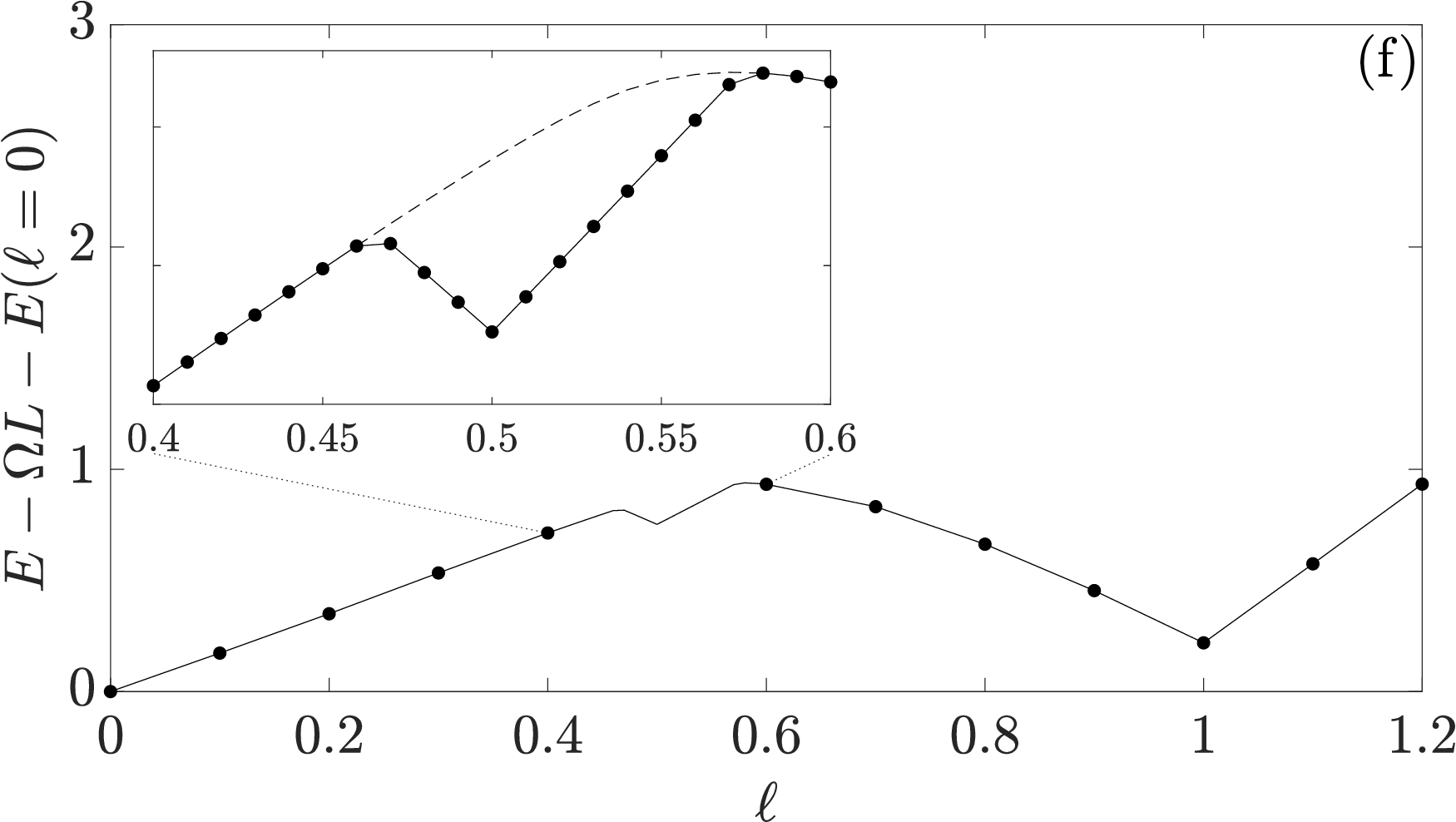}
\caption{(Cont.) (d) and (e) Same as (a)\textendash(c) but for $\ell = 0.57$ and $0.6$, respectively. (f) The corresponding dispersion relation in the rotating frame, with $\Omega = 0.03$. The unit of energy is $E_0$ and the unit of angular momentum is $\hbar$.}
\end{figure}

Most importantly, this simple semianalytic model shows that the crossover to the heterosymmetric state is associated with the value of $N \omega$, that is, with the confinement strength, with the heterosymmetric state being favored by stronger confinement. This behavior is in agreement with our numerical results. Let us consider the energy difference $E_\mathrm{h} - E_\mathrm{pl}$. In our model, for a given $N$, the contribution to $E_\mathrm{h} - E_\mathrm{pl}$ arising from the kinetic-plus-potential and contact energy terms increases linearly with $\omega$. At the same time, the contribution due to the logarithmic energy term decreases faster than linearly with $\omega$, due to the logarithmic factor multiplying the integrals $I_{2,\mathrm{pl}}$ and $I_{2,\mathrm{h}}$. For $\omega = \omega_c$, this decrease exactly counterbalances the increase of the rest of the energy terms, making the heterosymmetric state the yrast state beyond this line, where $E_\mathrm{h} < E_\mathrm{pl}$. We stress that both the decrease of the logarithmic energy term in the heterosymmetric state, as well as the corresponding increase of the rest of the terms, have also been observed in the numerical results.

From our model follows that increasing the value of $D$ moves the crossover towards higher values of $\omega$. This has also been confirmed by our numerical results. As we will see below, the same holds for an increasing imbalance ratio $\delta N / N$. It is worth noting that both these cases are associated with an increase of the contact energy term.

\subsubsection{Numerical results}

To illustrate our results regarding the heterosymmetric states in more detail, let us focus on a specific case, with $N = 200$ and $\omega = 0.05$. This value of the trapping frequency places the droplet above the critical line of Fig.\,2. In Fig.\,3 we present selected numerical results for the yrast states of the rotating droplet, for various angular momentum values close to $L = N/2$, or equivalently, $\ell = 0.5$. For $\ell = 0.4$ we see that the droplet exists in the phase-locked state, with a vortex having penetrated the surface of both components [Fig.\,3(a)]. Between $\ell = 0.46$ and $0.47$ there exists a discontinuous transition from a phase-locked to a heterosymmetric state, which is shown in Fig.\,3(b). Here, a vortex has entered only component $\downarrow$, along with a partially filled core, in the same position, in component $\uparrow$. Note that the phase of the order parameter of component $\uparrow$ displays no discontinuity line, confirming that the density dip is not a true vortex.

As $\ell$ increases, the off-center vortex moves towards the center of the trap. For $\ell = 0.5$ the vortex (and the partially filled core) naturally settle at the origin, in an axially symmetric configuration, as shown in Fig.\,3(c). As the angular momentum increases further, the droplet develops (octupole) surface waves in the heterosymmetric state, as shown in Fig.\,3(d). Finally, between $\ell = 0.57$ and $0.58$, there is again a discontinuous transition, this time from the heterosymmetric state back to the phase-locked state [Fig.\,3(e)]. For $\ell \geq 0.58$, the yrast states of the droplet turn out to be the same as those predicted by the single-order-parameter model \cite{NKO1}, which means that no other heterosymmetric states appear as yrast states in this case.

In Fig.\,3(f) we also present the dispersion relation in the rotating frame, i.e., $E_\mathrm{rot}(\ell)-E(\ell=0)$ where $E_\mathrm{rot}(\ell)=E(\ell)-\Omega L$, for a chosen value of the angular velocity $\Omega=0.03$. Plotting the dispersion relation in the rotating frame instead of the laboratory frame serves to make the structure of the curve more clearly visible. The inset focuses on the range of angular momentum values that is relevant to the emergence of the heterosymmetric states. We see that the curve displays a kink for $\ell = 0.5$, that is, a structure with a discontinuity in the slope of the curve. This structure corresponds to the axially symmetric heterosymmetric vortex state, and is similar to the structure that appears in various superfluid systems, when the angular momentum becomes equal to the number of atoms of the condensate (or component) that holds a vortex. Such a kink also appears in the present droplet system for $\ell = 1$, as seen in Fig.\,3(f), and corresponds to a phase-locked, axially symmetric single vortex in both components \cite{NKO1}.

The curve demonstrates that the rotating droplet initially lowers its energy by entering the heterosymmetric excitation mode, rather than remaining in the phase-locked mode (which is displayed by the dashed line in the inset). However, the emergence of surface waves for $\ell > 0.5$ increases the energy of the heterosymmetric state. In particular, the energy as a function of $\ell$ increases more rapidly in that excitation mode, compared to the phase-locked state with an off-center vortex, and eventually the phase-locked state is favored again. In both transitions, to and from the heterosymmetric mode, there is a level crossing in the dispersion relation, indicative of the discontinuous nature of the transitions.

It is instructive to note how the angular momentum is distributed between the two components of the droplet in the various yrast states we have uncovered. This is presented in Fig.\,4, where we have focused on the same range of $\ell$ values as in the inset of Fig.\,3(f). In the phase-locked states the angular momentum is shared equally between the two components. When the yrast state transitions to the heterosymmetric mode, the angular momentum is carried predominantly by component $\downarrow$, that is, the component that carries the vortex. More specifically, for $\ell < 0.5$, component $\uparrow$ carries a minor fraction of the total angular momentum, which arises from the deviation from axial symmetry (due to the off-center partially filled core). Then, for $\ell = 0.5$, we have $L_\uparrow = 0$ and $L_\downarrow = N_\downarrow = N/2$, that is, the entirety of the angular momentum is carried by component $\downarrow$. For $\ell > 0.5$, the emergence of surface waves imparts angular momentum to both components, with the majority still carried by component $\downarrow$. Finally, when the droplet transitions back to the phase-locked mode, the angular momentum is again shared equally.

\begin{figure}[t]
\centering
\includegraphics[width=\columnwidth]{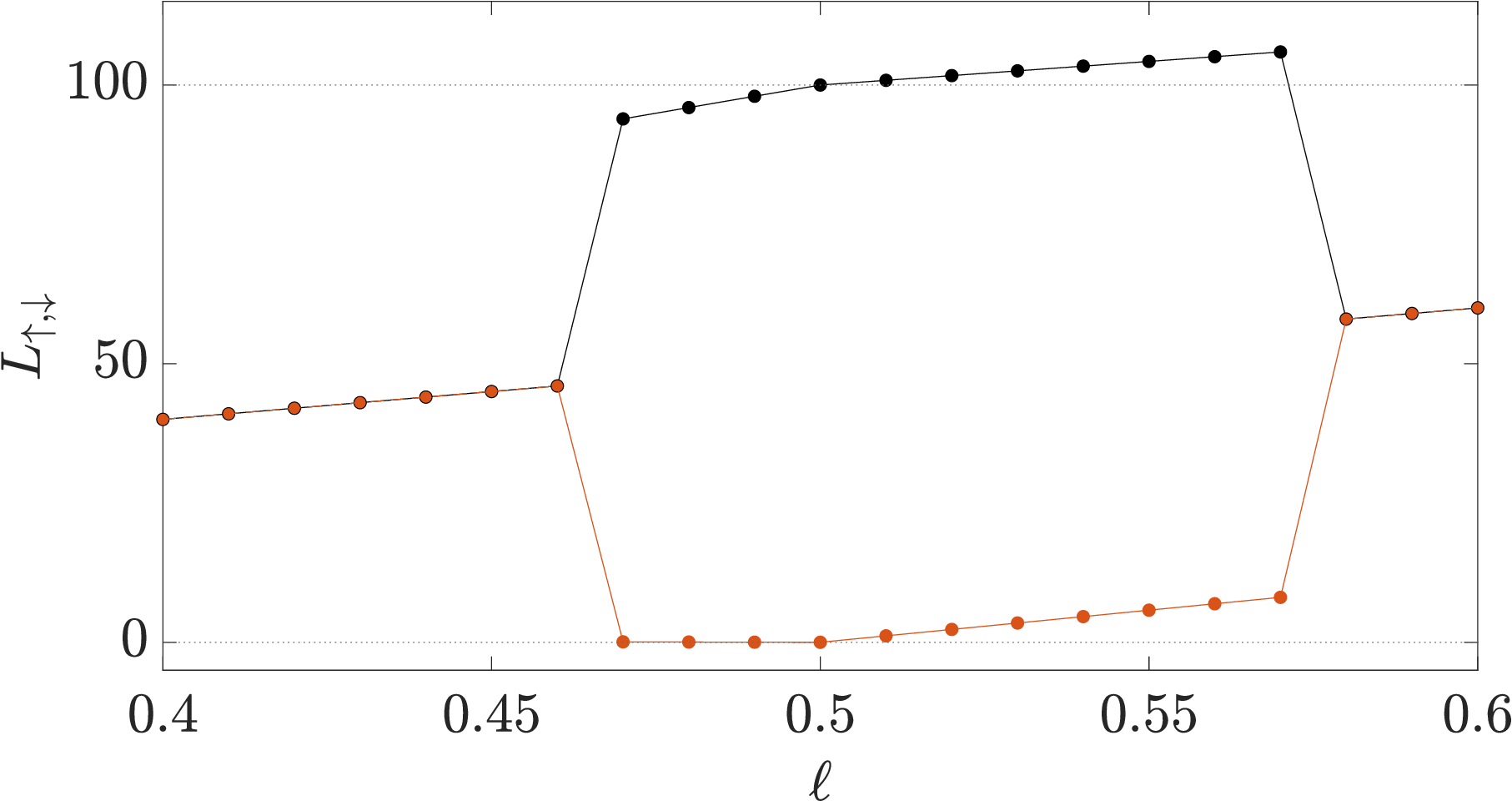}
\caption{The angular momenta of component $\uparrow$ (orange line) and component $\downarrow$ (black line) in the yrast state, as a function of the total angular momentum per particle $\ell$. Here $N = 200$, $\delta N = 0$, $\omega = 0.05$, and $D = 25$. The upper dotted horizontal line corresponds to $L = N_\uparrow = N_\downarrow = N/2$. The unit of angular momentum is $\hbar$.}
\end{figure}

\begin{figure}
\centering
\includegraphics[width=\columnwidth]{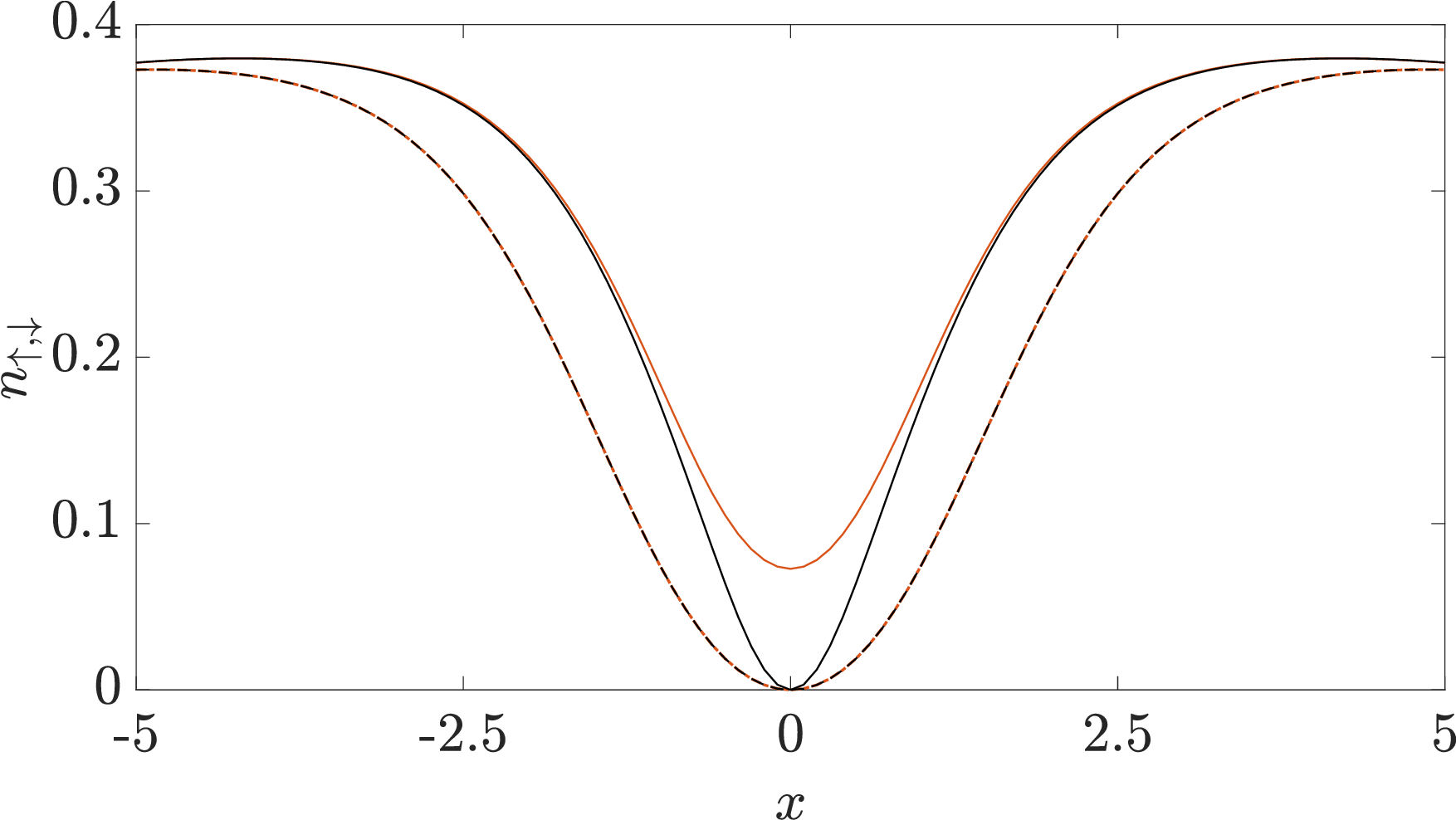}
\caption{The vortex profiles for the $L = N_\downarrow$ state (solid lines) and the $L = N$ state (dashed lines), with $N = 200$, $\delta N = 0$, $\omega = 0.05$, and $D = 25$. Black lines refer to component $\downarrow$ and orange lines refer to component $\uparrow$. The unit of density is $\Psi_0^2$ and the unit of length is $x_0$.}
\end{figure}

It is important to note that, in the balanced case we treat here, the vortex appearing in component $\downarrow$ instead of component $\uparrow$ is a consequence of the specific initial conditions we chose for the order parameters. Indeed, we have just as straightforwardly produced heterosymmetric states where the vortex is located in component $\uparrow$ instead, by interchanging the trial order parameters between the two components. These states have the same energy as the states we present here, i.e., the heterosymmetric states in a balanced droplet are doubly degenerate for each angular momentum value, with each pair related by interchanging the location of the vortex and the partially filled core between the components. This double degeneracy is a manifestation of the underlying $\mathbb{Z}_2$ symmetry, which is spontaneously broken \cite{SSB}. In practice, the presence of even slight deviations from our assumption of a balanced system, that is, a population asymmetry, will lift this degeneracy via explicit breaking of the $\mathbb{Z}_2$ symmetry, as we demonstrate in the following subsection. We stress that such deviations are effectively unavoidable in an experimentally realistic situation. We also note that in a balanced droplet with an interaction asymmetry the populations $N_{\uparrow,\downarrow}$ are different, that is, the system is not $\mathbb{Z}_2$-symmetric under the exchange of the $\uparrow$ and $\downarrow$ atoms, so this degeneracy does not arise \cite{Poparic}.

Another interesting aspect of the heterosymmetric states is that the radial width of the vortex core exhibits a marked decrease, compared to that of a vortex existing in both components. This is illustrated in Fig.\,5, where we plot the vortex profiles for $L=N_\downarrow$ and $L=N$. According to our numerical results, the radius of a vortex existing in only one component, calculated at the density half maximum, is approximately $35\%$ smaller than the width of a vortex existing in both components. From a physical point of view, the non-zero density at the partially filled core in component $\uparrow$ induces an attraction to the particles in component $\downarrow$, which in effect decreases the vortex width. A similar decrease of the vortex radius in the heterosymmetric state has also been noted in the case of a heteronuclear quantum droplet \cite{Caldara}.

\subsection{Heterosymmetric vortex states of imbalanced droplets}

Let us now turn to the case of an imbalanced self-bound droplet, investigating the effects of a small population asymmetry between the two components. An obvious characteristic of the imbalanced droplet is that the two components will display a density difference between them. Regarding the heterosymmetric states in particular, an interesting feature of the imbalanced system is that there exist two different values of the angular momentum, $L = N_\downarrow$ and $L = N_\uparrow$, which are expected to support an axially symmetric, heterosymmetric state, lifting the aforementioned degeneracy that appears in the balanced system. Here, we repeat that $\uparrow$ and $\downarrow$ refer to the majority and minority component, respectively. In other words, the presence of a nonzero population imbalance explicitly breaks the $\mathbb{Z}_2$ symmetry, as the system is no longer invariant under the exchange of the $\uparrow$ and $\downarrow$ atoms. As a result, the double degeneracy of the heterosymmetric yrast states is lifted. Indeed, we have found that when the angular momentum increases towards $N_\downarrow$ the droplet transitions to a state where the vorticity is carried by component $\downarrow$, and when it approaches $N_\uparrow$ it transitions to a state where the vorticity is carried by component $\uparrow$, instead.

\begin{figure}
\centering
\includegraphics[width=\columnwidth]{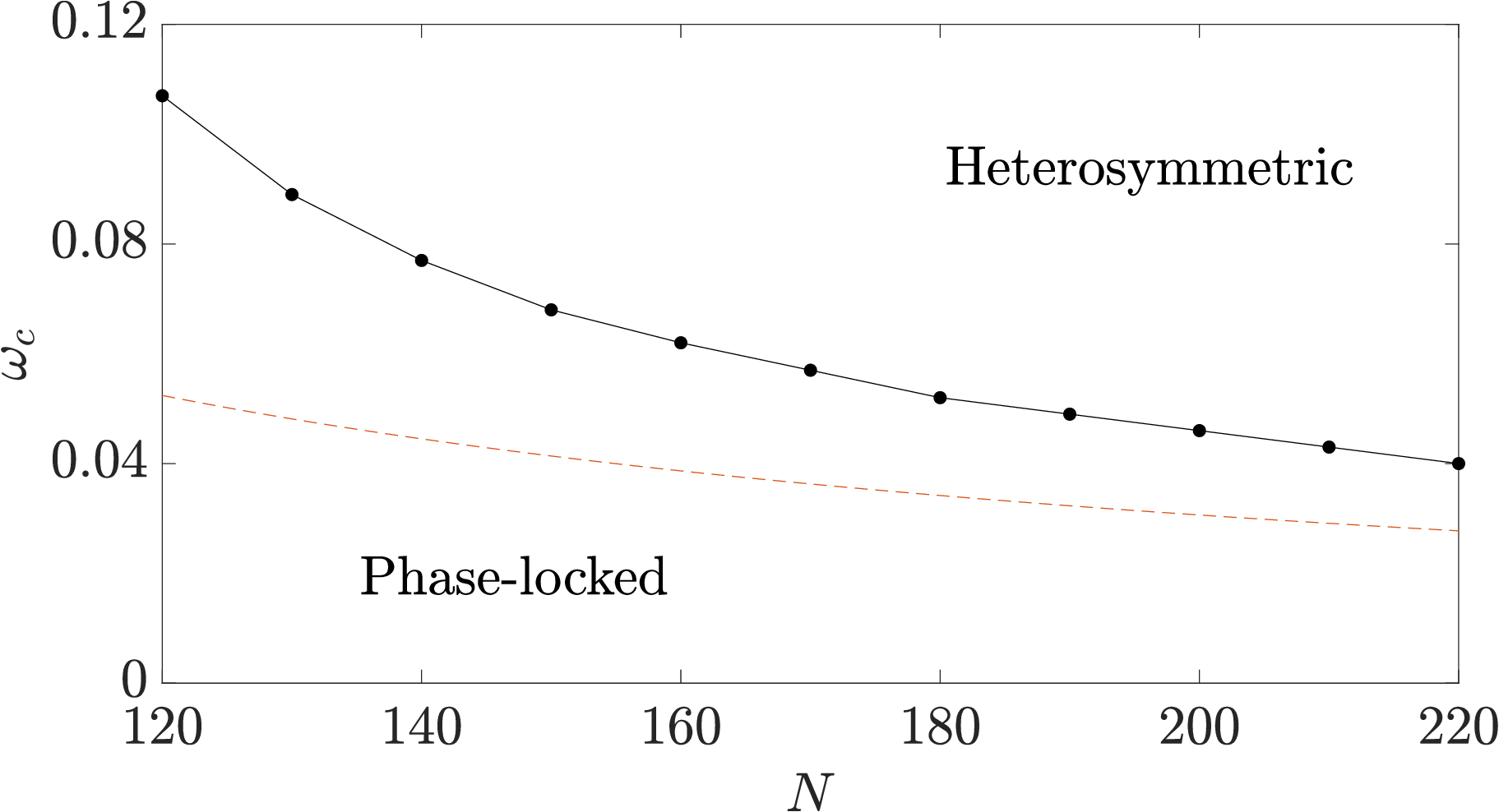}
\caption{Solid line, with data points: the critical value of the trapping frequency $\omega$ for which a heterosymmetric state, with a vortex only in the $\downarrow$ component, appears as the yrast state, as a function of the total number of atoms $N$. Here $L = N_\mathrm{\downarrow}$, $\delta N/N = 0.025$, and $D=25$. Dashed line: the semianalytic result we have derived for the crossover.}
\end{figure}

The requirement of strong enough confinement for the heterosymmetric states to appear as yrast states also holds for the imbalanced case. In particular, we have used the semianalytic model presented above to calculate the critical value of $\omega$ for the crossover to the heterosymmetric state, as a function of $N$, for an imbalance ratio $\delta N/N = 0.025$. We note that for this value of $\delta N/N$ we have found that the coefficients of Eq.\,(\ref{heterosymmetric_up_order_parameter}) take the values $d_0 = 0.73$ and $d_1 = 0.27$, which are slightly different than the balanced case. We present the result of the calculation, along the fully numerical results, in Fig.\,6. As in the balanced case, the accuracy of our semianalytic model improves with increasing $N$. In addition, we see that the presence of the population imbalance moves the crossover towards slightly higher values of $\omega$, although this is more pronounced in smaller $N$. Below the critical line, as in the balanced case, the droplet exists in a phase-locked state, but with a density difference between the two components.

\begin{figure}
\centering
\includegraphics[width=\columnwidth]{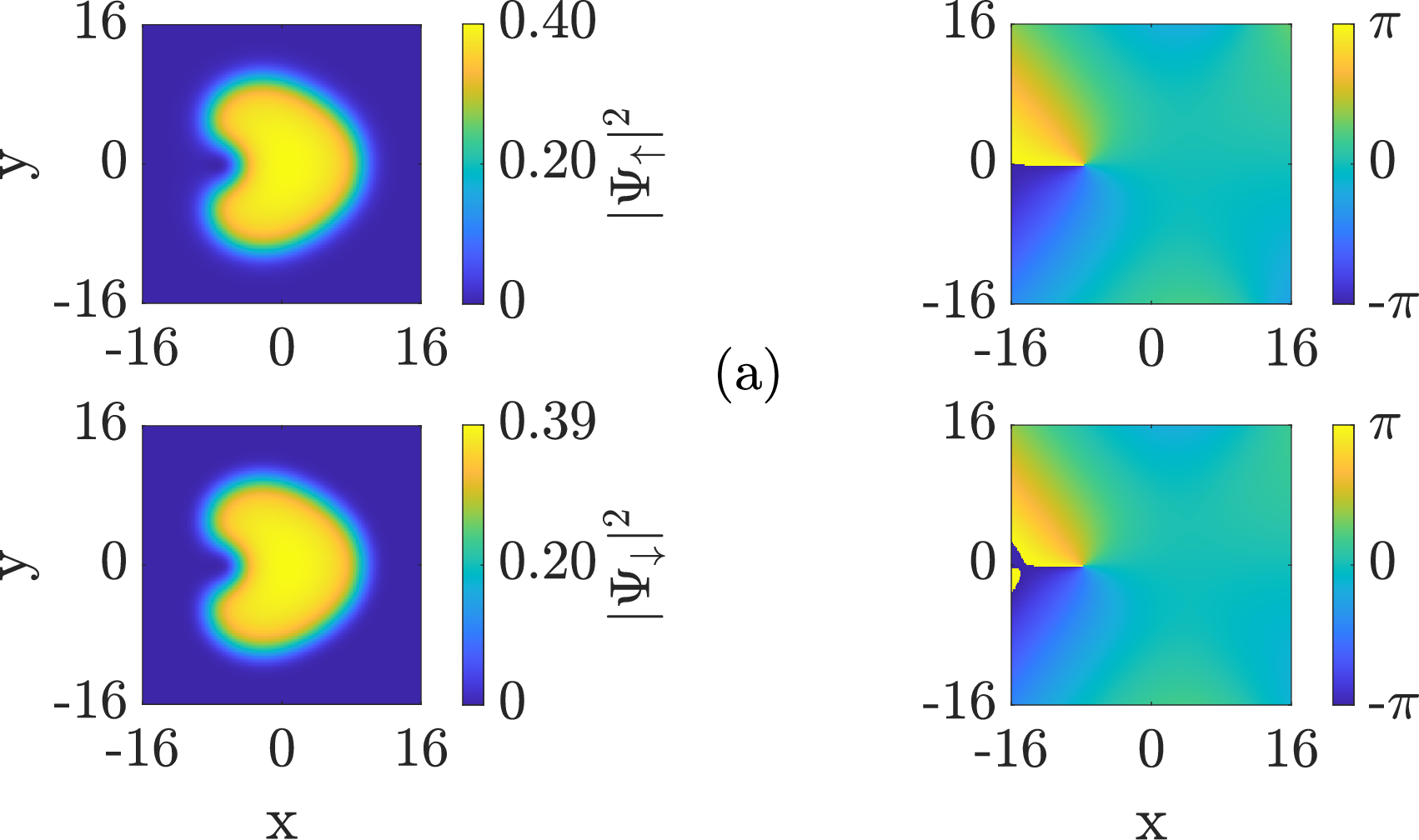}\\
\vspace{0.3\baselineskip}
\includegraphics[width=\columnwidth]{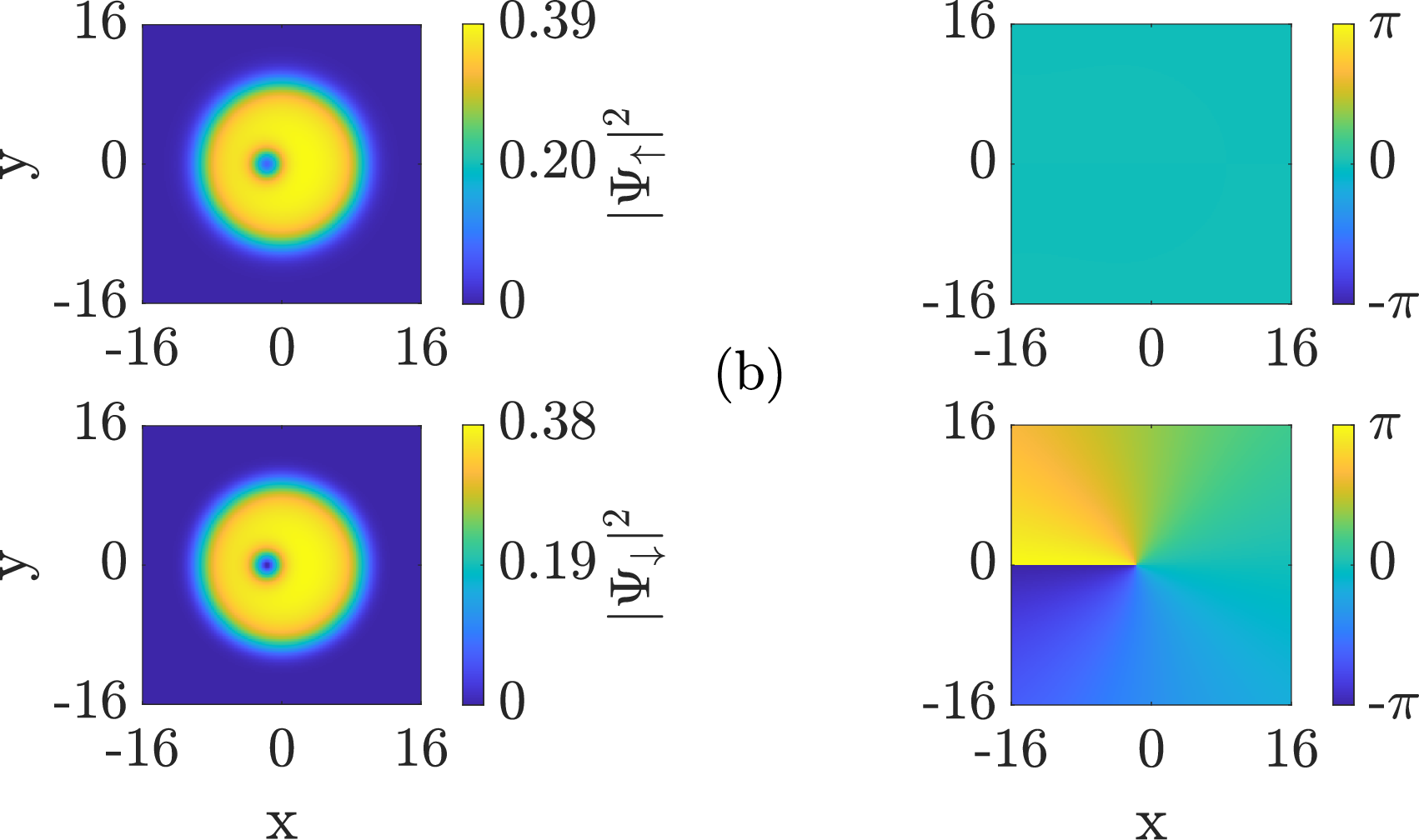}\\
\vspace{0.3\baselineskip}
\includegraphics[width=\columnwidth]{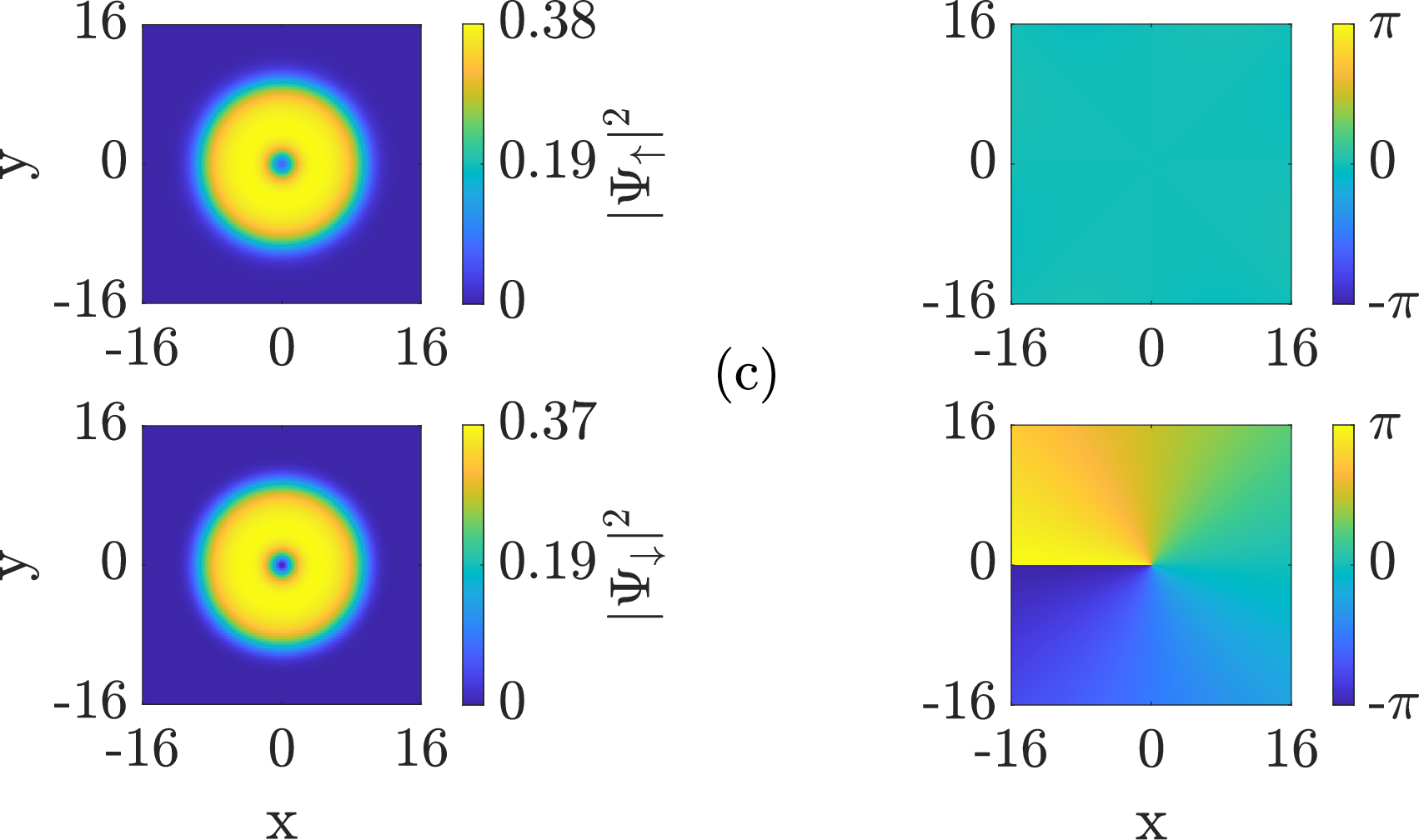}\\
\vspace{0.3\baselineskip}
\includegraphics[width=\columnwidth]{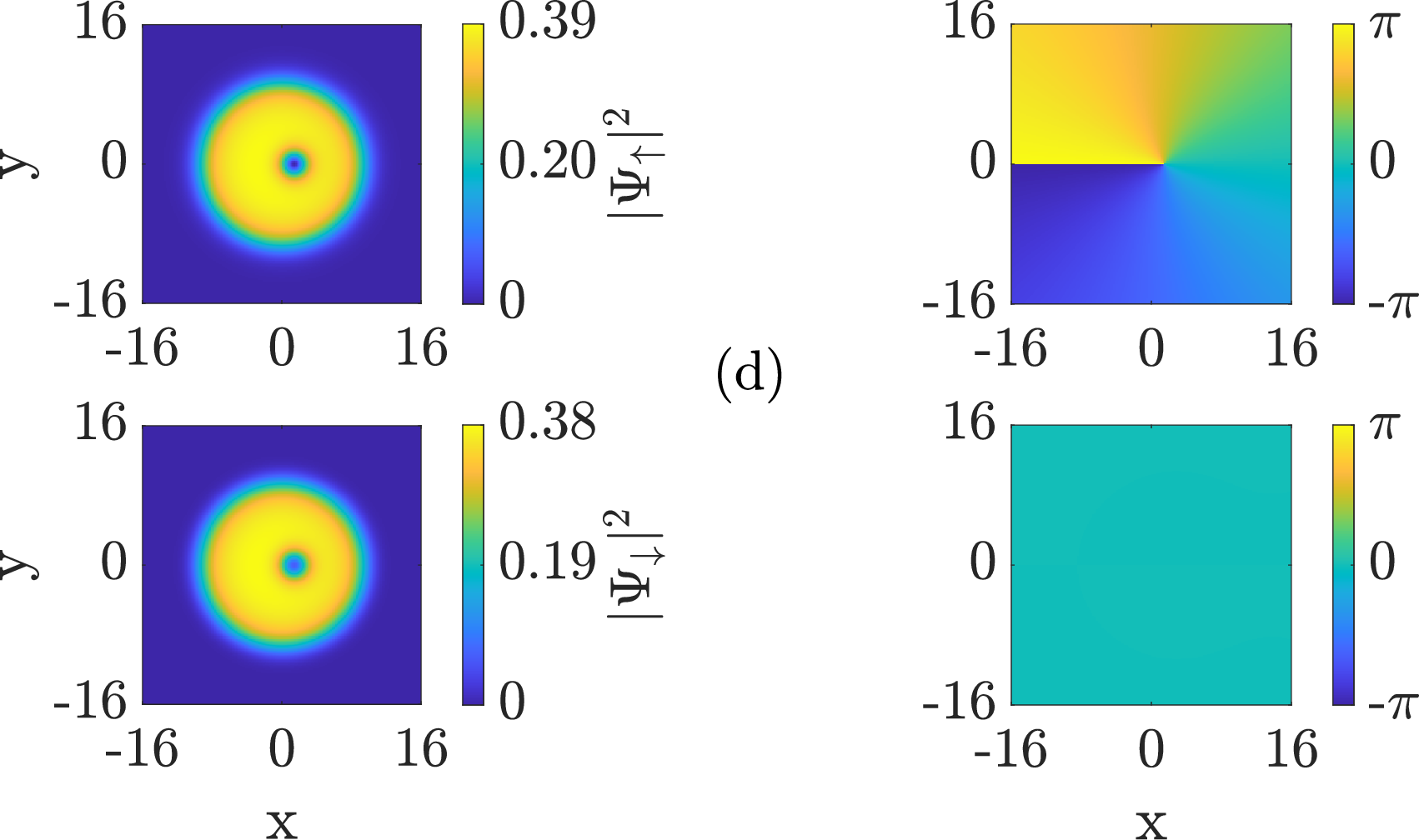}
\caption{(a)\textendash(d) The densities (left column, in units of $\Psi_0^2$) and the phases (right column) of the droplet order parameters, in the yrast state, for $N = 200$, $\omega = 0.05$, $\delta N=5$, $D=25$, and (a) $\ell = 0.4$, (b) $\ell = 0.47$, (c) $\ell = 0.4875$, and (d) $\ell = 0.5$. The unit of length is $x_0$.}
\addtocounter{figure}{-1}
\end{figure}

\begin{figure}
\centering
\includegraphics[width=\columnwidth]{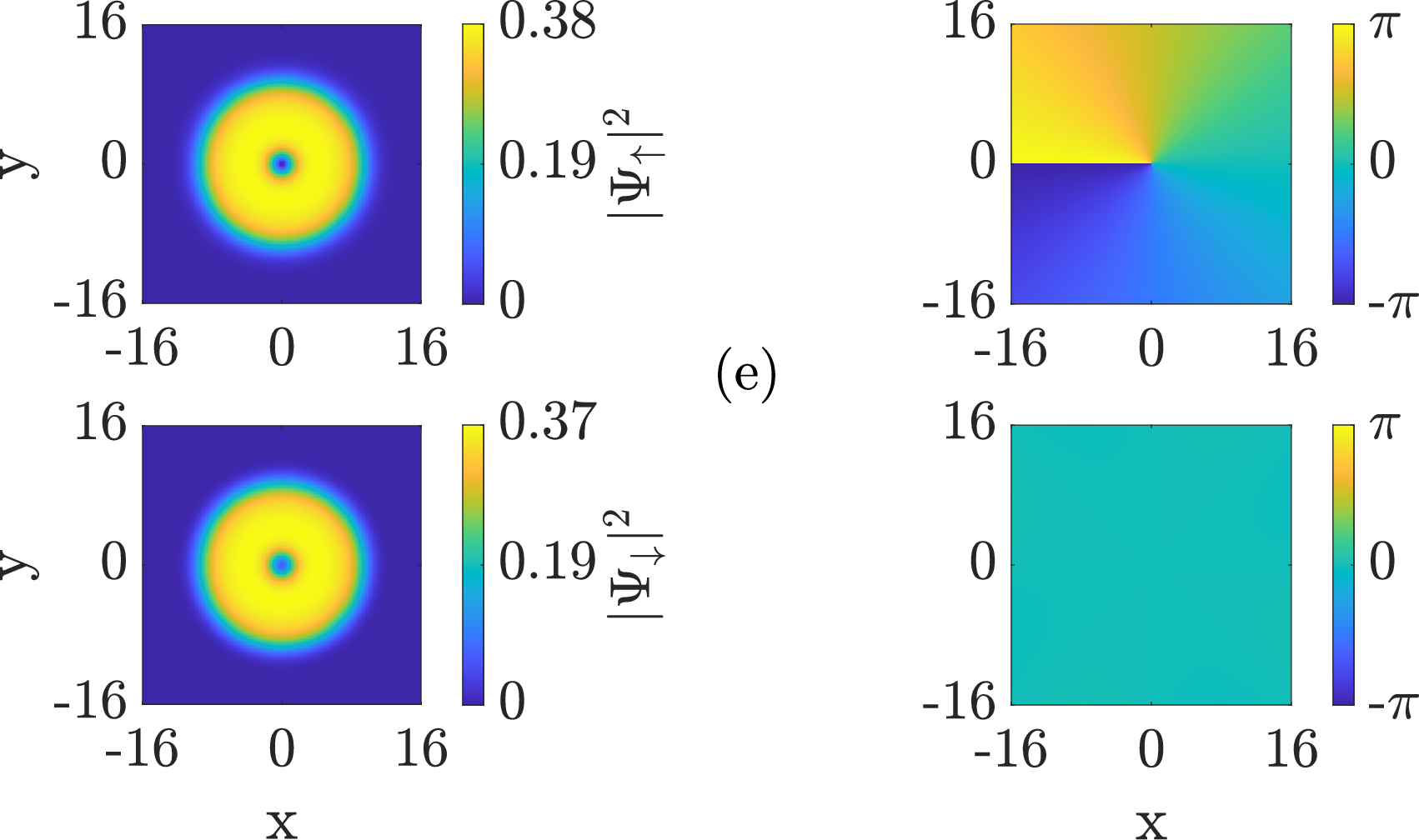}\\
\vspace{0.3\baselineskip}
\includegraphics[width=\columnwidth]{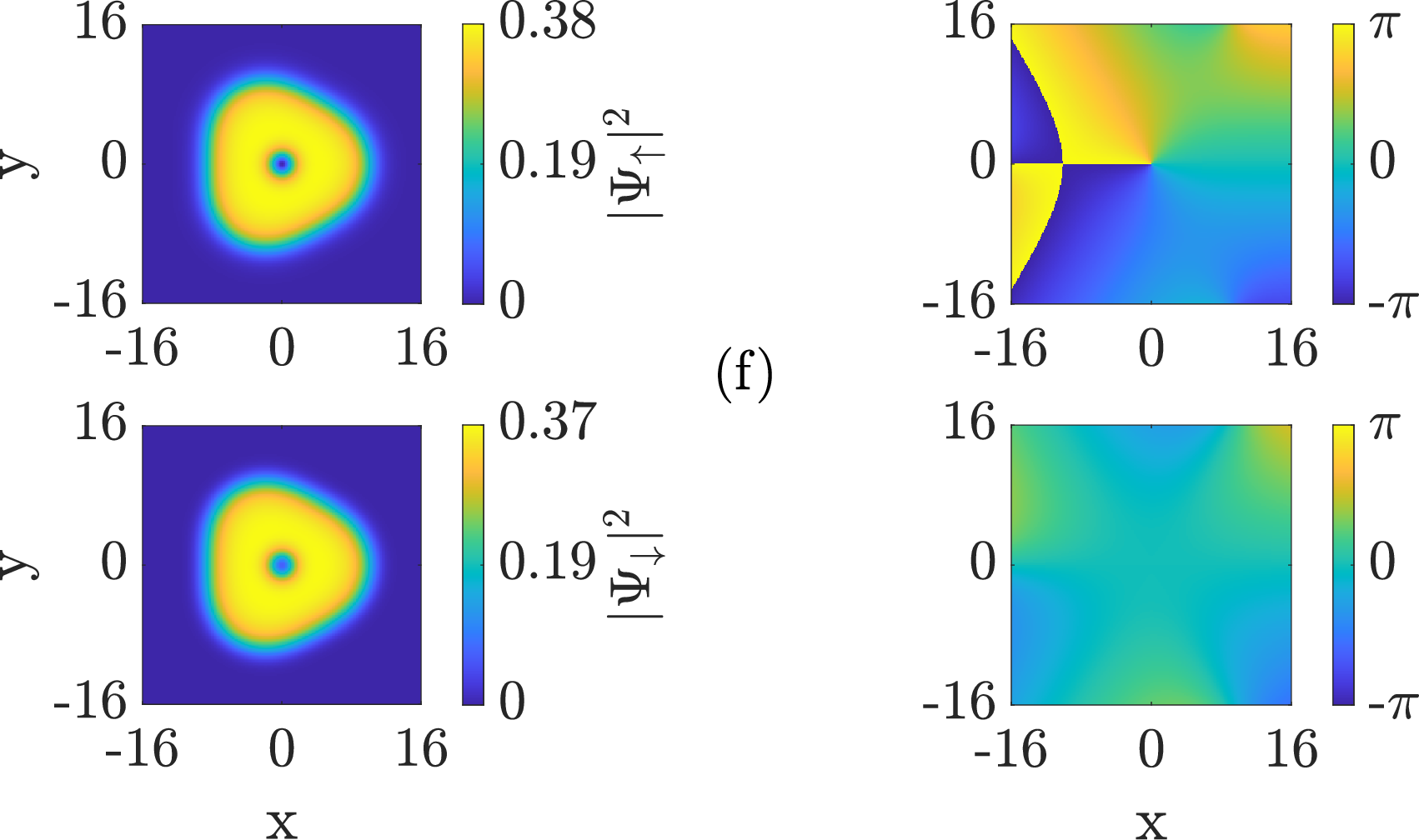}\\
\vspace{0.3\baselineskip}
\includegraphics[width=\columnwidth]{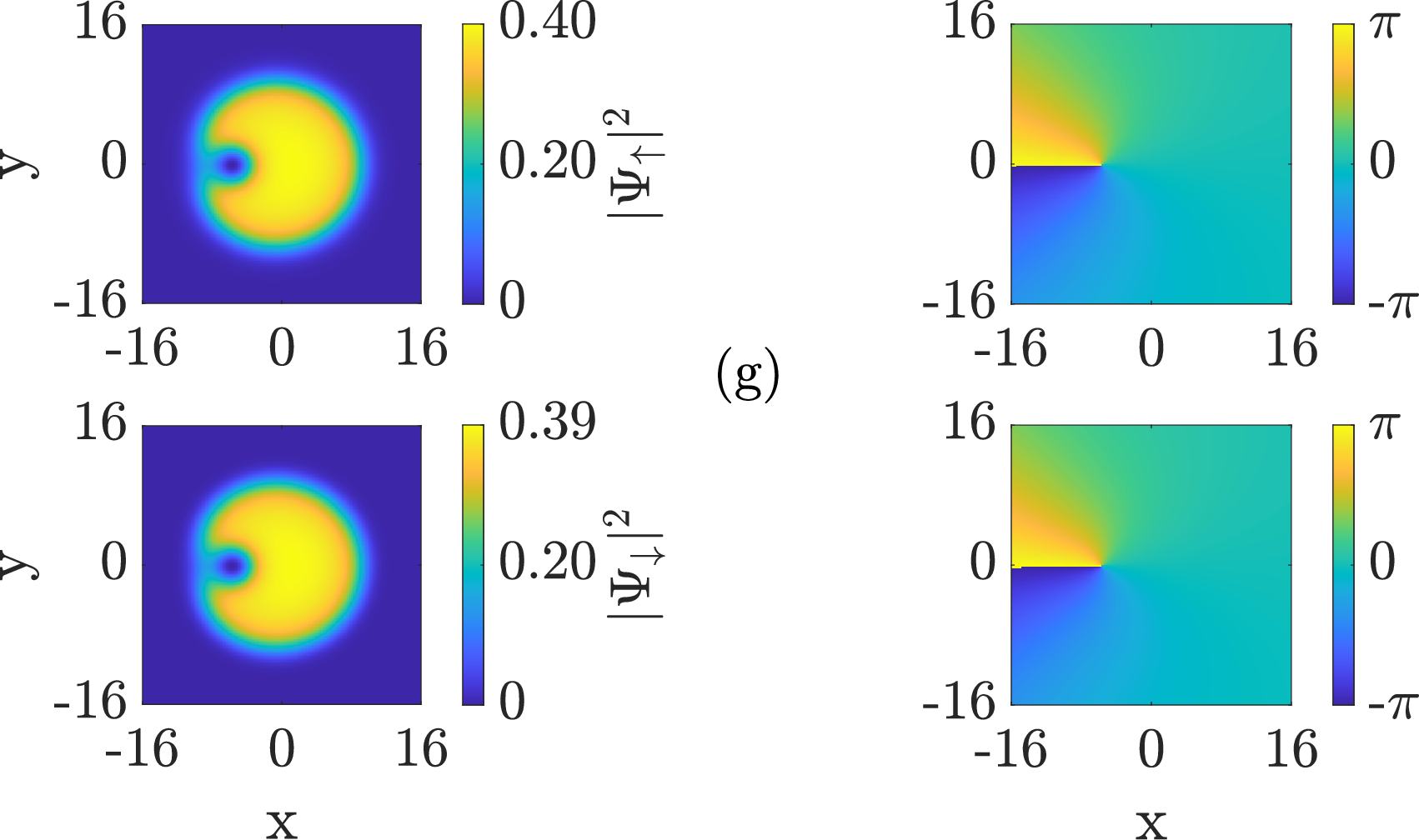}\\
\vspace{0.3\baselineskip}
\includegraphics[width=\columnwidth]{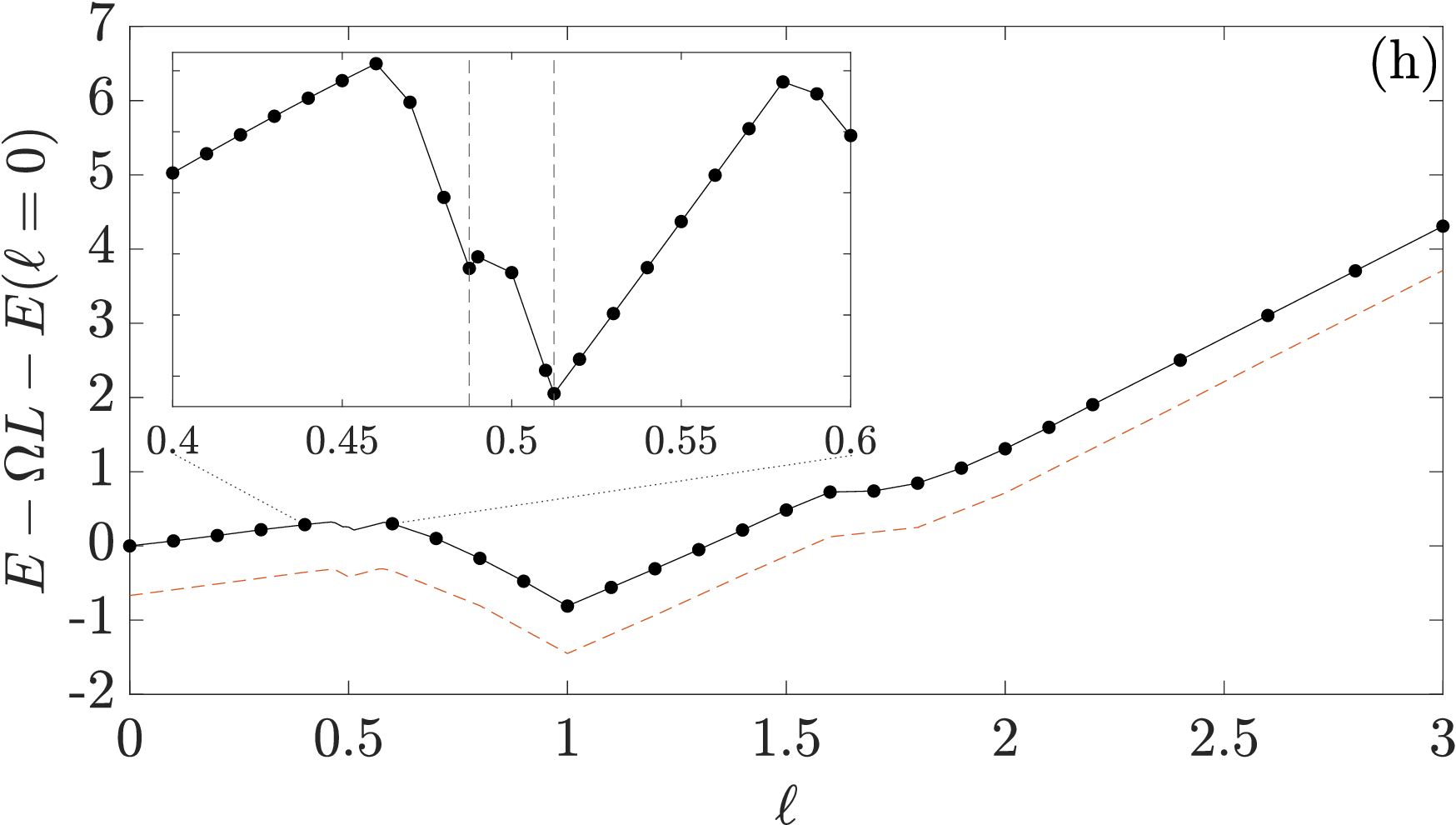}\\
\caption{(Cont.) (e)\textendash(g) Same as (a)\textendash(d) but for $\ell = 0.5125$, $0.57$, and $0.6$, respectively. (h) Solid line, with data points: The corresponding dispersion relation in the rotating frame, with $\Omega = 0.035$. The dashed vertical lines mark the values $L = N_\downarrow$ and $N_\uparrow$. Dashed line: The dispersion relation of the balanced droplet ($\delta N = 0$). The unit of energy is $E_0$ and the unit of angular momentum is $\hbar$.}
\end{figure}

For the presentation of specific numerical results, we focus on the same values for the total number of atoms and the trapping frequency as in the balanced case, that is, $N = 200$ and $\omega = 0.05$, while we additionally set $\delta N = 5$. In total, we have varied the angular momentum per particle from $\ell = 0$ up to $3$, calculating the yrast state for each value. Selected results from these calculations are presented in Fig.\,7, where we again focus mainly on the region of heterosymmetric excitation. For $\ell = 0.4$ [Fig.\,7(a)], the droplet exists in a phase-locked state, with a vortex penetrating both components. Here, we note the density difference between the two components, due to the population asymmetry. For $\ell = 0.47$ [Fig.\,7(b)] the yrast state turns to be a heterosymmetric state, with an off-center vortex in component $\downarrow$. As in the balanced case, the transition from the phase-locked to the heterosymmetric state, which is located between $\ell=0.46$ and $0.47$, is first-order, i.e., the slope of the dispersion relation exhibits a discontinuity, associated with a level crossing. For $\ell=0.4875$, that is $L = N_\downarrow$, the vortex in component $\downarrow$ settles at the origin [Fig.\,7(c)]. Between $\ell=0.49$ and $0.5$ there exists another discontinuous transition, from a heterosymmetric state with a vortex in the component $\downarrow$ to another heterosymmetric state, with an off-center vortex in component $\uparrow$, as displayed in Fig.\,7(d). For $\ell=0.5125$, that is $L = N_\uparrow$, this heterosymmetric state again turns axially symmetric [Fig.\,7(e)]. Then, for $\ell > 0.5125$, it also develops surface waves at both components, as shown in Fig.\,7(f) for $ell=0.57$. Finally, another discontinuous transition appears between $\ell=0.58$ and $0.59$, and the yrast state turns to be a phase-locked one, as shown in Fig.\,7(g) for $\ell=0.6$. Beyond this value of the angular momentum, the vortex in both components continues approaching the origin, until, as expected, it settles at the center of the droplet for $\ell = 1$.

In general, for an imbalanced rotating droplet, we have found that, for sufficiently high values of the angular momentum (in our specific case $\ell \geq 0.59$), the two components are excited in the same manner, and the picture is qualitatively similar to the balanced case, apart from the noted density difference among the two components. In the case presented here in particular, for $\ell > 1$ a second singly-quantized vortex eventually enters both components of the droplet. At this point, the droplet becomes saturated with vortices, in the sense that it is not energetically favorable to host additional vortices as the angular momentum increases further. Then, for $\ell > 2.4$, the additional angular momentum is carried via center-of-mass excitation of the state that hosts two vortices. This ``mixed" mode of excitation constitutes a novel aspect in the rotational response of self-bound quantum droplets, and has been studied in detail for the balanced case \cite{NKO1,NKO3}.

In Fig.\,7(h) we present the corresponding dispersion relation in the rotating frame, for a chosen value of the angular velocity $\Omega=0.035$. We also plot the dispersion relation of the balanced droplet, measured against the nonrotating ground state energy $E(\ell=0)$ of the imbalanced droplet. The inset of Fig.\,3(h) focuses in the region $0.4 \leq \ell \leq 0.6$. We see that the dispersion relation here displays two separate kinks (points of slope discontinuity), for $L=N_\downarrow$ and $N_\uparrow$, which correspond to the two axially symmetric heterosymmetric states discussed above. This is the most striking difference between the balanced and the imbalanced case regarding the heterosymmetric rotating states. Essentially, when $\delta N$ takes a nonzero value, the single kink that appears for $L = N_\downarrow = N_\uparrow = N/2$ (with $\delta N = 0$), separates into two kinks, as shown on the inset, which signifies that the degeneracy of the two heterosymmetric states is effectively lifted by the population asymmetry. As in the balanced droplet, another kink appears for $\ell = 1$, which is actually the global minimum of the curve for this particular value of $\Omega$. Finally, for $\ell > 2.4$ the dispersion relation turns linear, with a slope equal to $\omega=0.05$. This is a characteristic of the center-of-mass motion of the (vortex carrying) droplet \cite{NKO1}, as the nonlinear energy terms do not depend on $L$ for $\ell > 2.4$. Quantitatively, and as is apparent in Fig.\,7(h), the dispersion relation of the imbalanced droplet exhibits an energy increase compared to the balanced one, owed to the positive contribution of the contact energy term (which is identically zero for the phase-locked states in the balanced system).

\begin{figure}
\centering
\includegraphics[width=\columnwidth]{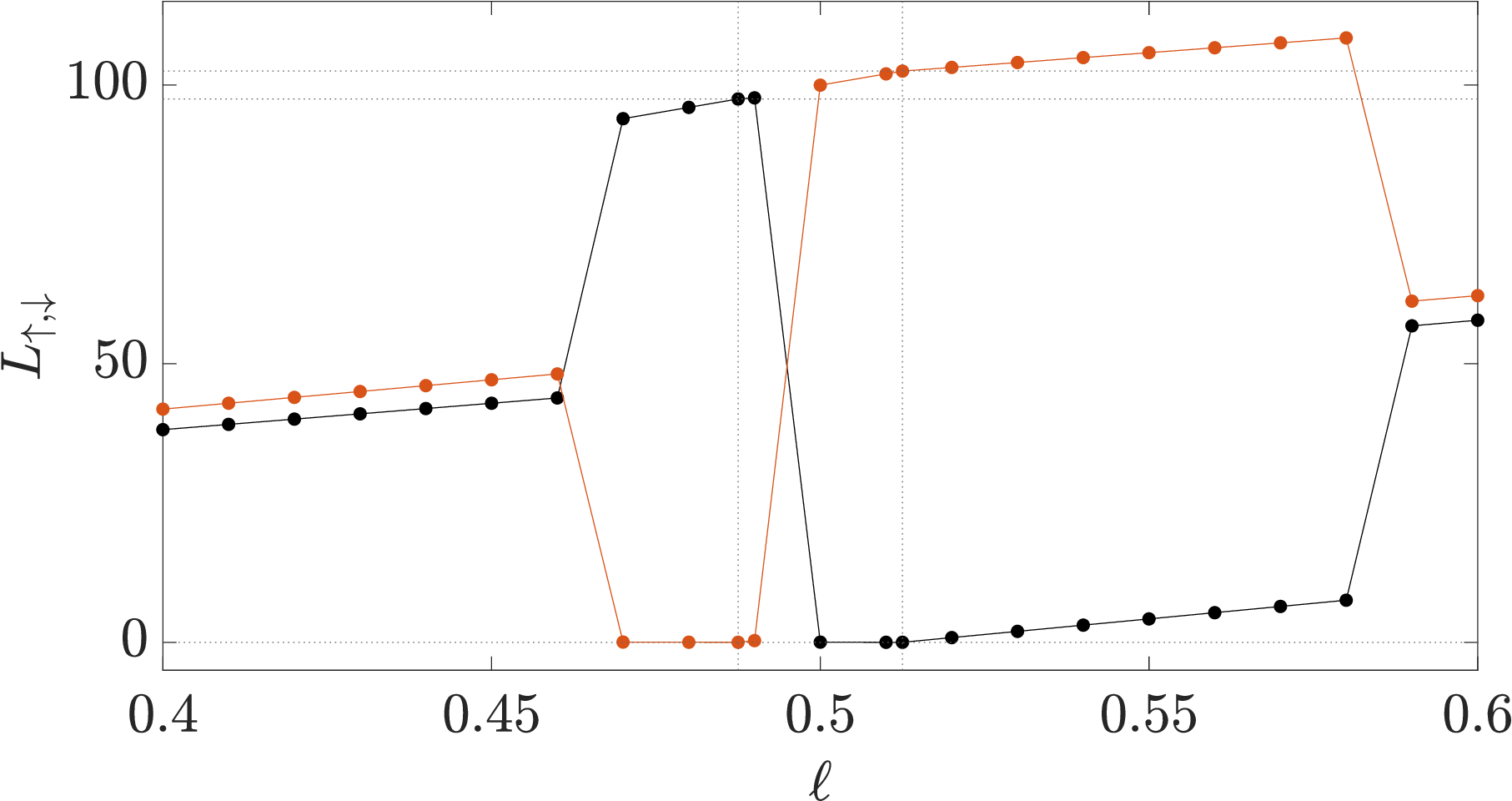}
\caption{The angular momenta of component $\uparrow$ (orange line) and component $\downarrow$ (black line) in the yrast state, as a function of the total angular momentum per particle $\ell$. Here $N = 200$, $\delta N = 5$, $\omega = 0.05$, and $D = 25$. The dotted vertical lines correspond to $L = N_\downarrow$ and $N_\uparrow$, as do the upper dotted horizontal lines. The unit of angular momentum is $\hbar$.}
\end{figure}

As in the balanced case, it is interesting to see how the angular momentum is distributed in the two components, in the region of heterosymmetric excitation. This is presented in Fig.\,8. We see that in the phase-locked states, the angular momentum is shared between the two components, albeit not equally, but rather proportionally to each population. When the droplet transitions to the heterosymmetric state with a vortex in component $\downarrow$, the angular momentum is carried predominantly by this component (in particular for $L = N_\downarrow$, $L_\downarrow = N_\downarrow$ and $L_\uparrow = 0$). The opposite happens when the droplet transitions to the heterosymmetric state with a vortex in component $\downarrow$. Here, the angular momentum is carried predominantly by component $\uparrow$, and in particular for $L = N_\uparrow$, we have $L_\downarrow = 0$ and $L_\uparrow = N_\uparrow$. The abrupt crossings of the angular momenta $L_{\uparrow,\downarrow}$ signify the discontinuous nature of these transitions. Finally, when the droplet transitions back to the phase-locked mode, the angular momentum is again shared proportionally.

\section{Yrast states of anharmonically confined quantum droplets}

Let us now turn to the case of anharmonic confinement of rotating quantum droplets. In the study of rotating condensates, it is well known that for a harmonic confining potential the rotational frequency of the trap $\Omega$ is limited by the trap frequency $\omega$, as the centrifugal potential scales quadratically with the radial distance. Therefore, when $\Omega > \omega$, the effective potential obtains an inverse-parabolic form, becoming unbounded from below. For a quantum droplet in particular, this rapid rotation regime is associated with center-of-mass motion of a vortex-carrying state \cite{NKO1,NKO3}. In contrast, an anharmonic confining potential effectively obtains a sombrero shape for $\Omega > \omega$, with its minima located in a circular region around the center of the trap, that is, it remains bounded regardless of the value of $\Omega$. This has important consequences for the rotational response of the condensate, especially in the rapid rotation regime. 

\begin{figure}
\centering
\includegraphics[width=\columnwidth]{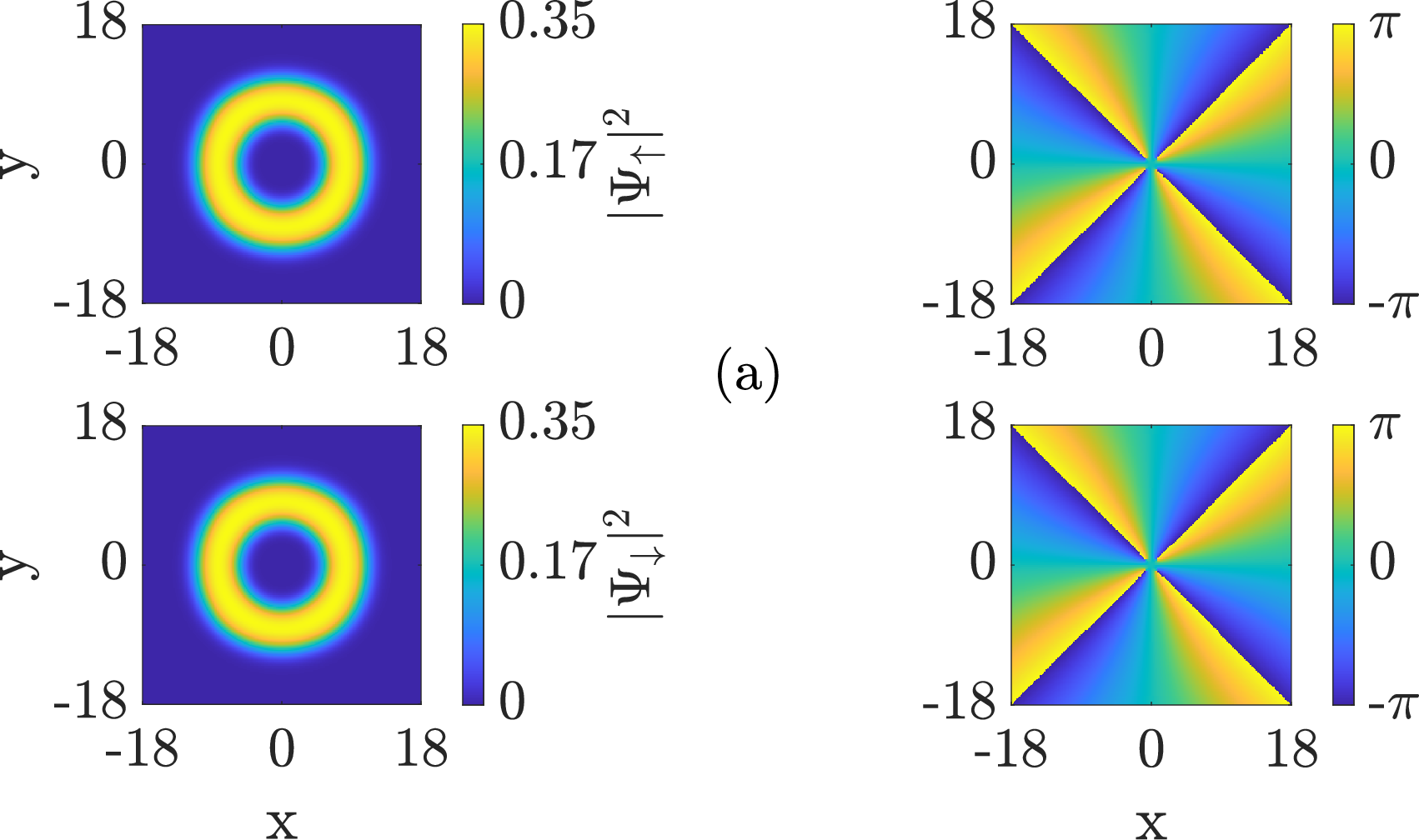}\\
\vspace{0.3\baselineskip}
\includegraphics[width=\columnwidth]{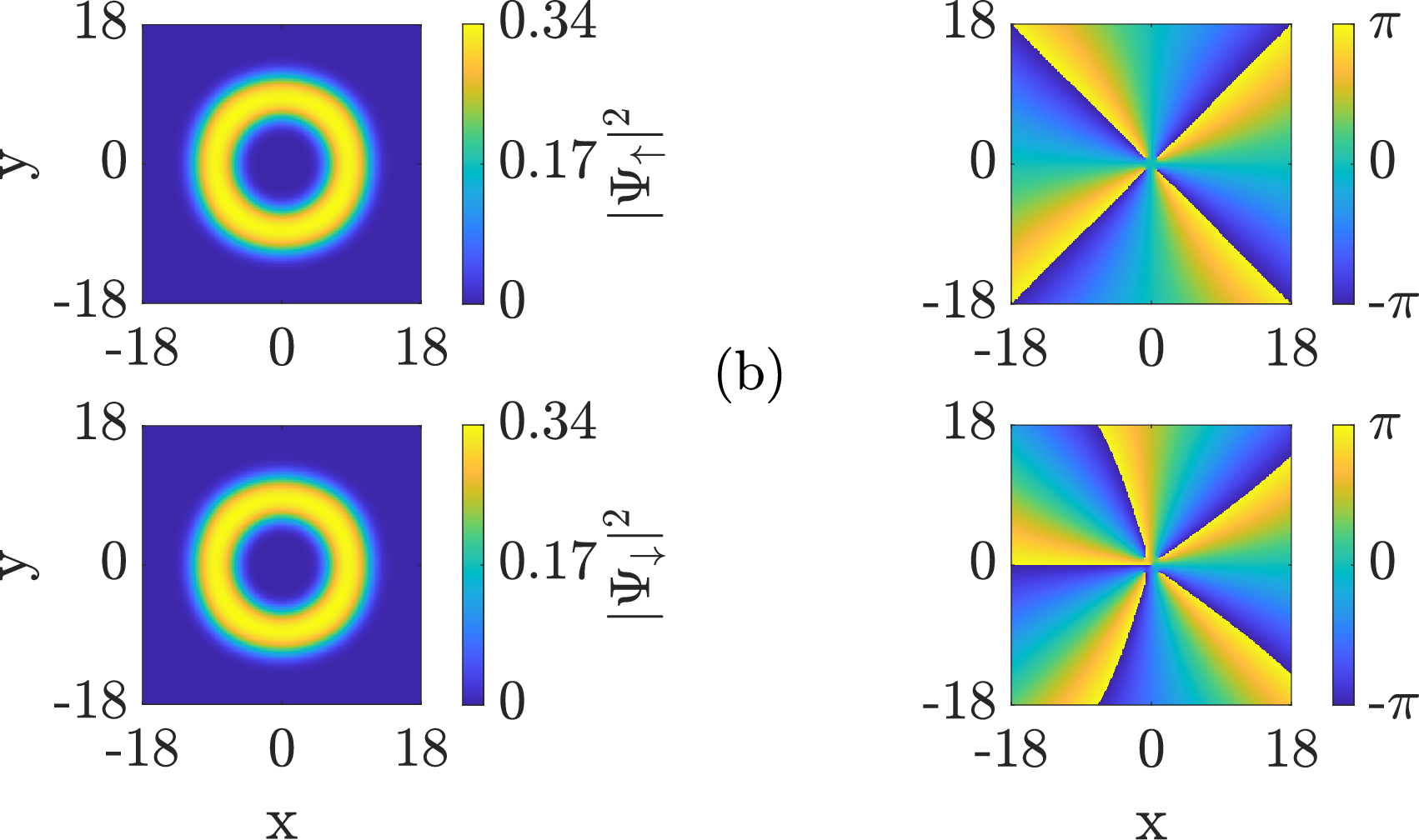}\\
\vspace{0.3\baselineskip}
\includegraphics[width=\columnwidth]{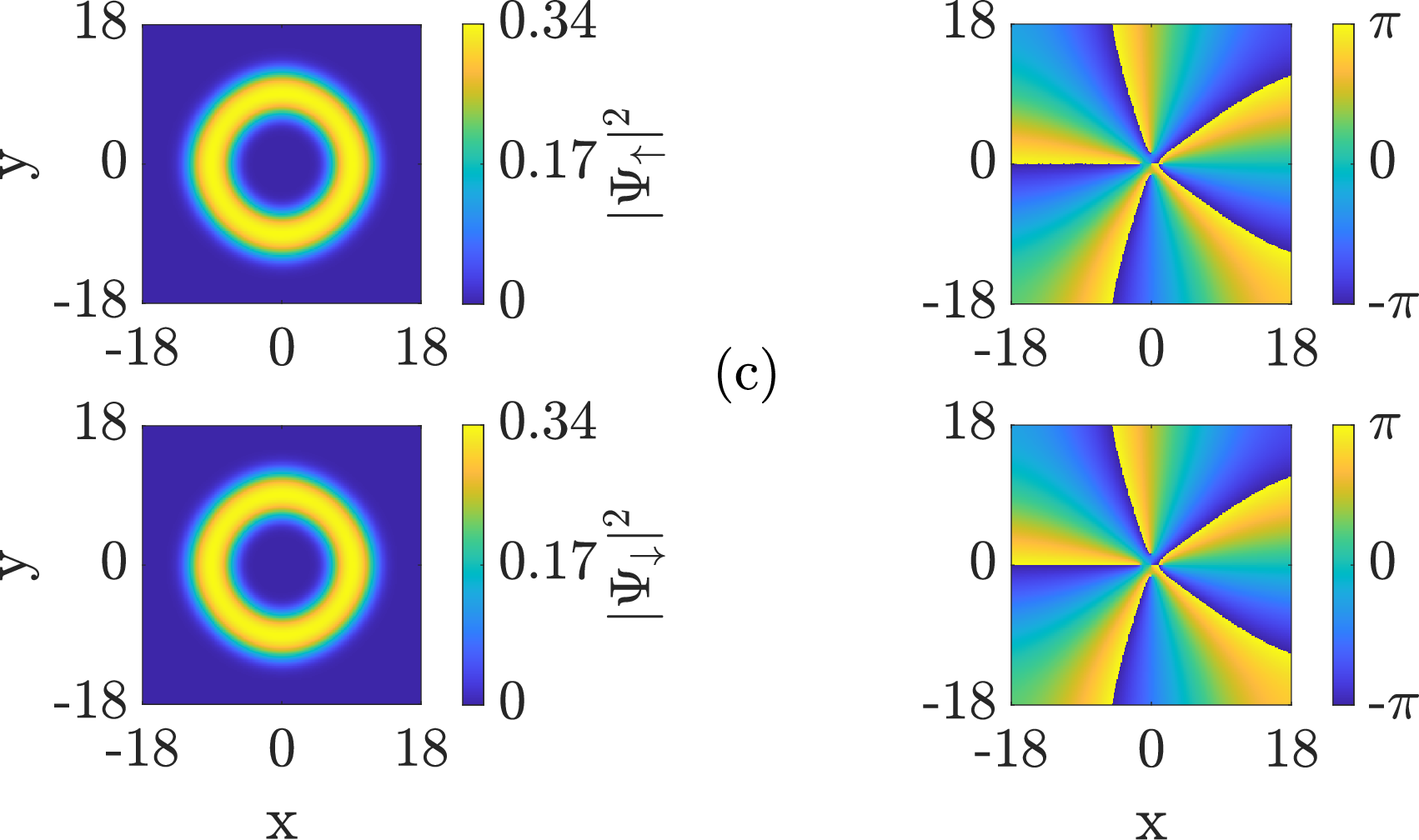}
\caption{(a)\textendash(c) The densities (left column, in units of $\Psi_0^2$) and the phases (right column) of the droplet order parameters, in the yrast state, for $N = 200$, $\omega = 0.05$, $\lambda = 0.05$, $\delta N = 0$, $D=25$, and (a) $\ell = 4$, (b) $\ell = 4.5$, and (c) $\ell = 5$. The unit of length is $x_0$.}
\addtocounter{figure}{-1}
\end{figure}

\begin{figure}[!ht]
\centering
\includegraphics[width=\columnwidth]{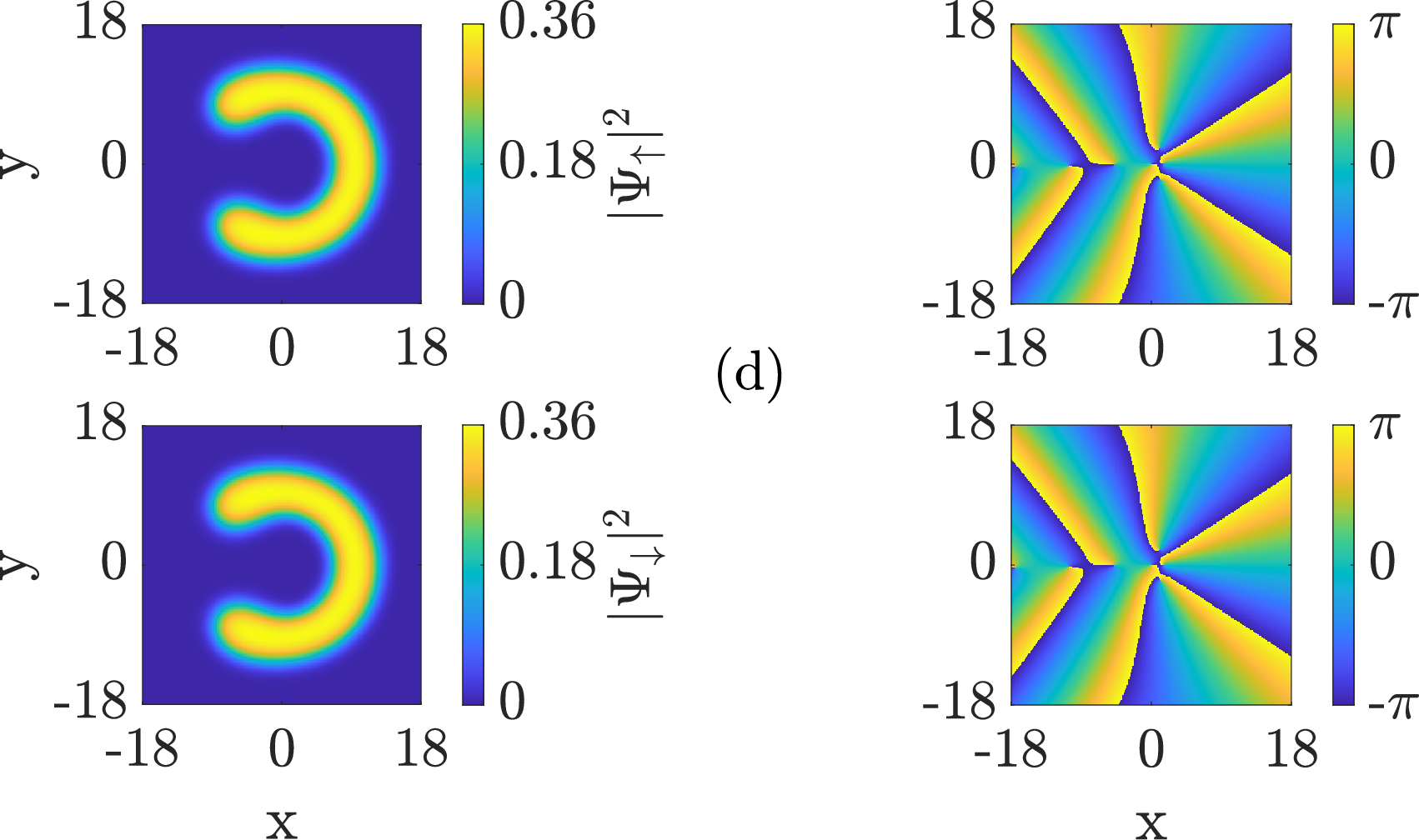}\\
\vspace{0.3\baselineskip}
\includegraphics[width=\columnwidth]{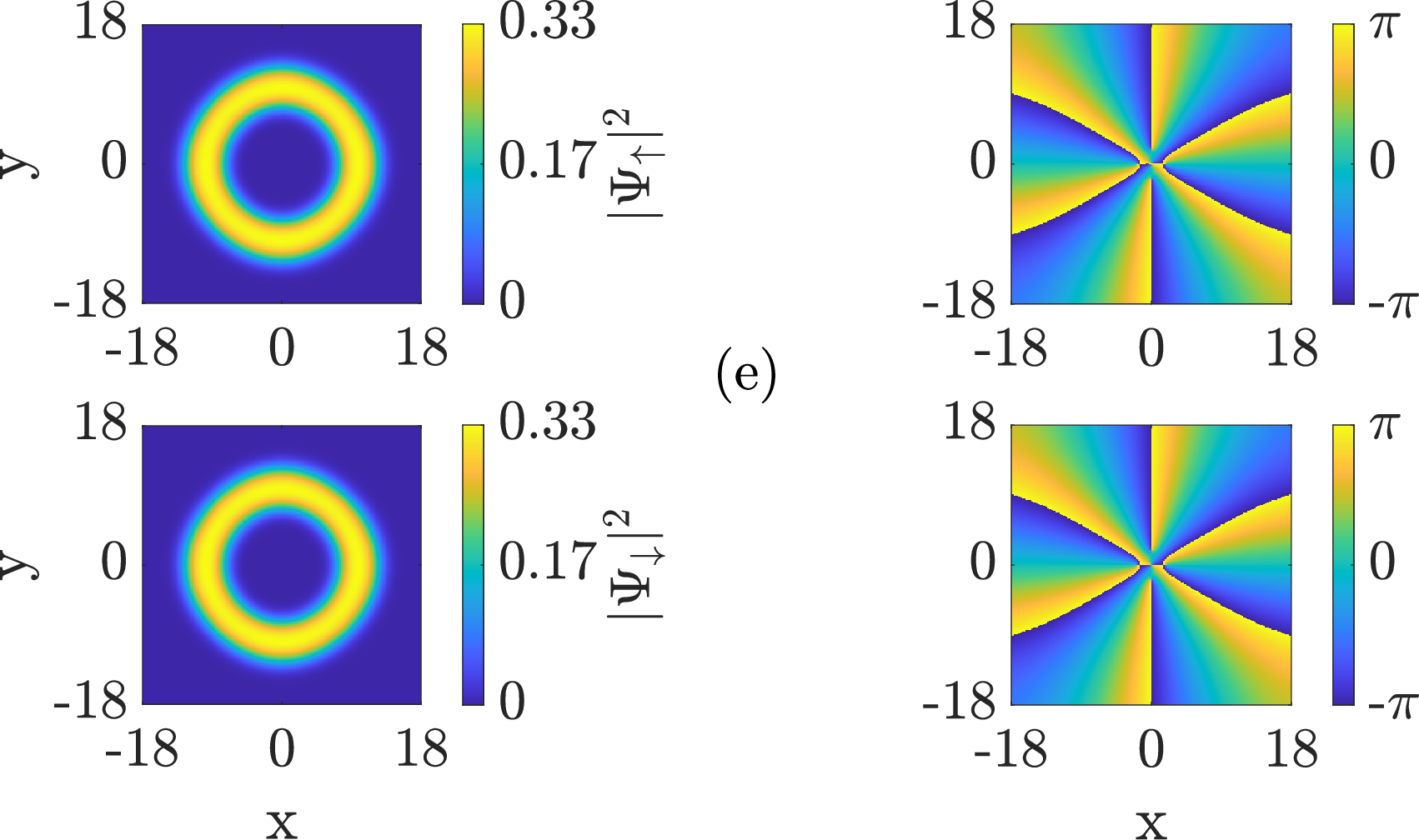}\\
\vspace{0.3\baselineskip}
\includegraphics[width=\columnwidth]{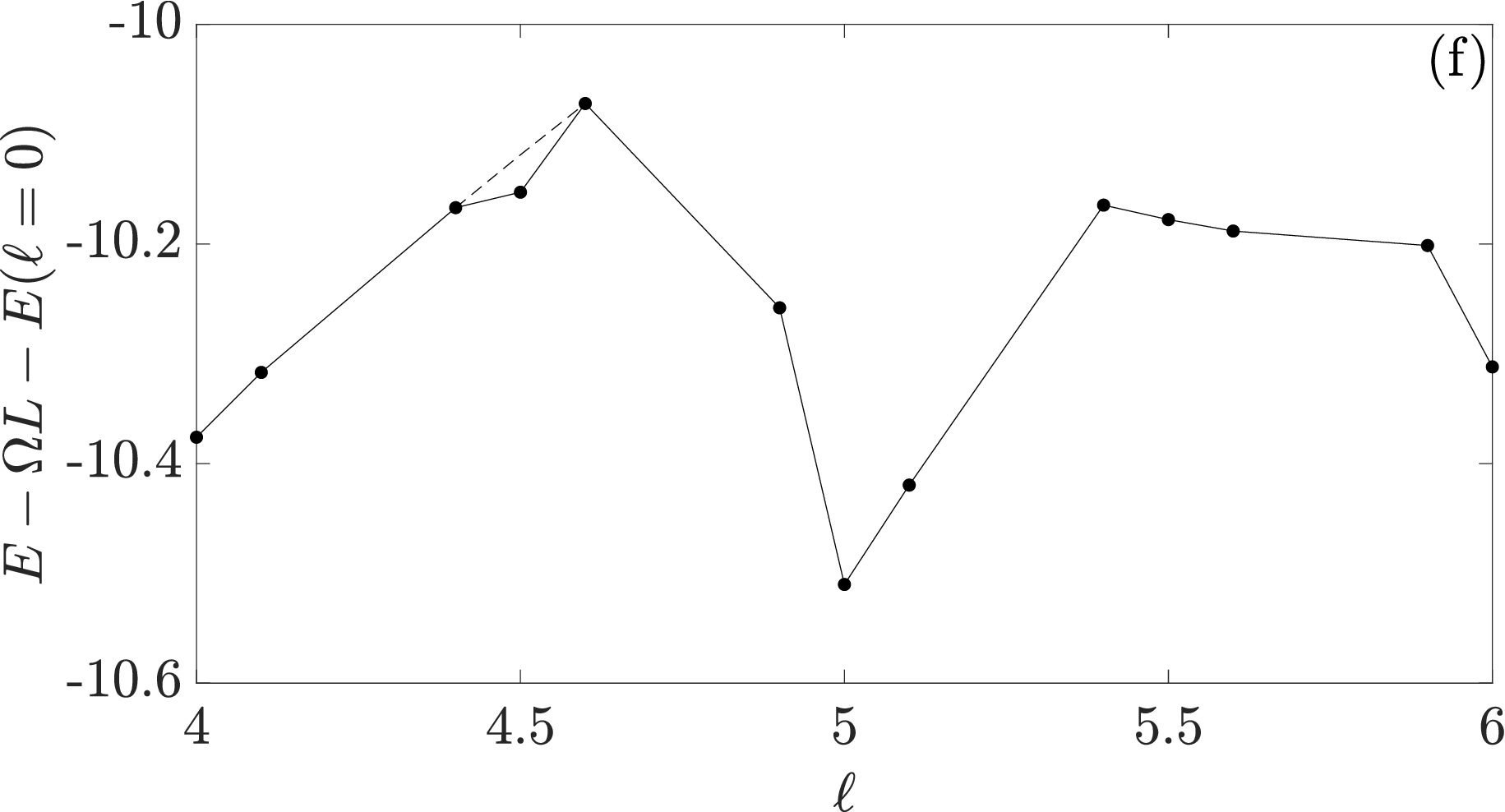}\\
\caption{(Cont.) (d), (e) Same as (a)\textendash(c) but for $\ell = 5.5$ and $6$, respectively. (f) The corresponding dispersion relation in the rotating frame, with $\Omega = 0.063$. The unit of energy is $E_0$ and the unit of angular momentum is $\hbar$.}
\end{figure}

One hallmark of anharmonic confinement in this context is the appearance of multiply quantized vortex states as the yrast state, for certain values of the angular momentum, facilitated by the sombrero shape of the effective potential. This is also the case for droplets in particular, for an appropriate set of parameter values \cite{NKO2}. In the light of the heterosymmetric states that we have uncovered in the case of harmonic confinement, a natural question here is whether a similar picture arises for an anharmonic potential, with heterosymmetric multiply quantized vortex states appearing as yrast states, which would be missed by the usually employed single-order-parameter model. To explore this possibility within the present model, we focus on an anharmonically confined droplet with $N=200$, $\omega=0.05$, $\lambda = 0.05$, and $D=25$, calculating the yrast state for various values of the angular momentum per particle $\ell$. This particular set of values is useful in directly comparing our present results with those of the single-order-parameter model of Ref.\,\cite{NKO2}. We also consider both cases of a balanced and an imbalanced droplet. 

\subsection{Heterosymmetric multiply quantized vortex states of balanced droplets}

Initially, we have set $\delta N = 0$, focusing on the balanced case. In total, we have varied the angular momentum per particle up to $\ell = 7$. Naturally, for the heterosymmetric multiply quantized vortex states that we seek, half-integer values of $\ell$ are of particular interest. For such values, a possible rotating state is a multiply quantized vortex in each component, but with the quanta of circulation (i.e., the winding number) differing by one unit between the components. In general, we have used a step of $0.5$ for the $\ell$ values in the calculations, decreasing it to $0.1$ when more detail was required.

In particular, the yrast state we have found for $\ell = 0.5$ is a heterosymmetric state with a vortex in component $\downarrow$, while component $\uparrow$ does not carry vorticity, but rather a partially filled core. This state is essentially the same as the one we identified and analyzed for the harmonic potential, with only a slight increase in the droplet density (due to the addition of the quartic term in the potential, which strengthens the confinement). Then, for $0.5 < \ell < 3$, we have produced a series of phase-locked yrast states, holding up to three singly quantized vortices in both components. In this range of $\ell$ values, the results of the present calculation are identical to those of the single-order-parameter model \cite{NKO2}. For $\ell = 3$, the yrast state is a phase-locked, triply quantized vortex state. Moving further, $\ell = 3.5$ constitutes a candidate value for a heterosymmetric multiply quantized vortex state. However, the yrast state produced by our calculation is a phase-locked state with three singly quantized vortices in an asymmetric configuration, as in the single-order-parameter model \cite{NKO2}. 

The picture becomes more interesting for $\ell > 4$. In Fig.\,9 we focus on the results in the range $4 \leq \ell \leq 6$. For $\ell = 4$, the yrast state is a phase-locked, quadruply quantized vortex state, as shown in Fig.\,9(a). Then, for $\ell = 4.5$, the yrast state turns to be a heterosymmetric multiply quantized vortex state, with winding numbers equal to $4$ in component $\uparrow$ and $5$ in component $\downarrow$, as is apparent in the phase plots of Fig.\,9(b). Contrary to the phase-locked states, in which the angular momentum is shared equally between the two components, here it is distributed according to the carried vorticity: we have $L_\uparrow = 4N_\uparrow = 2N$ and $L_\downarrow = 5N_\downarrow = 2.5N$, given that $N_\uparrow = N_\downarrow = N/2$. Naturally, this state is missed by the single-order-parameter model, which predicts a phase-locked quadruply quantized vortex that deviates from axial symmetry \cite{NKO2}. For $\ell = 5$, the yrast state is a phase-locked, multiply quantized vortex state, with a winding number equal to $5$ (in both components), as presented in Fig.\,9(c). For $\ell = 5.5$, we have another candidate for a heterosymmetric state. Rather, the calculation has produced a phase-locked localized yrast state which breaks the axial symmetry, as shown in Fig.\,9(d). Then, for $\ell = 6$, the yrast state is again a phase-locked vortex state, with a winding number equal to $6$ in both components [Fig.\,9(e)]. Finally, for $6 < \ell \leq 7$, the yrast state has been found to be a phase-locked localized state, similar to the one that appears for $\ell = 5.5$. 

Figure 9(f) presents the corresponding dispersion relation in the rotating frame, for an angular velocity $\Omega = 0.063 > \omega$ and focused on the range $4 \leq \ell \leq 6$. For each of the integer values of $\ell$, which in this range correspond to phase-locked multiply quantized vortex states, there is a familiar kink/slope discontinuity. In the plot, this is more apparent for $\ell = 5$, which is the global minimum of the curve for this value of $\Omega$. Another such kink, albeit shallower, appears for $\ell = 4.5$ and corresponds to the heterosymmetric state. The dashed line connecting $\ell = 4.4$ to $4.6$ is the energy of the phase-locked state. Again, we see that the rotating droplet lowers its energy by entering the heterosymmetric state (via a level crossing corresponding to a discontinuous transition). Conversely, no kink appears for $\ell = 5.5$, where the yrast state is a phase-locked state. 

In total, we have found that the present model, which treats the two order parameters separately, generally reproduces the results of the single-order-parameter model, with the important exceptions of the heterosymmetric states that appear for $\ell = 0.5$ and $\ell = 4.5$. The latter in particular is a manifestation of the anharmonicity of the confining potential. We have found that the requirement of strong confinement for the appearance of heterosymmetric yrast states, as established for the harmonic potential, also applies for the anharmonic case we treat here. For example, for a decreased value of the trapping frequency, $\omega = 0.04$, and the rest of the parameter values unchanged, the heterosymmetric state with $\ell = 4.5$ does not appear as an yrast state, as a phase-locked state has lower energy. Conversely, for $\omega = 0.04$, the heterosymmetric state with $\ell = 0.5$ remains the yrast state, even though for a purely harmonic potential with the same frequency it is not the yrast state. We therefore deduce that the introduction of anharmonicity in the confining potential, which in effect strengthens the confinement, has the effect of slightly lowering the value of $\omega_c$, which is required for the crossover to the heterosymmetric state with a vortex only in one component. This effect may also be accounted for in the semianalytic model we have developed, by treating the quartic part of the potential as a perturbation.

The discussion regarding the double degeneracy of the heterosymmetric yrast states due to the spontaneous breaking of the $\mathbb{Z}_2$ symmetry, as laid out in the previous section, also applies for any possible multiply quantized vortex state. In particular, the heterosymmetric state we have uncovered for $\ell = 4.5$ has the same energy as a state where the circulations are interchanged between the two components. Again, and as we show below, a nonzero population asymmetry lifts this degeneracy by explicitly breaking the $\mathbb{Z}_2$ symmetry.

It is interesting to note that we have not uncovered any heterosymmetric yrast state with the vorticities of the two components differing by more than one unit. From our calculations of such states, we have noted that they are associated with a significant increase of the kinetic energy, which precludes them from appearing as yrast states, at least for the parameter values that we have explored.

\begin{figure}
\centering
\includegraphics[width=\columnwidth]{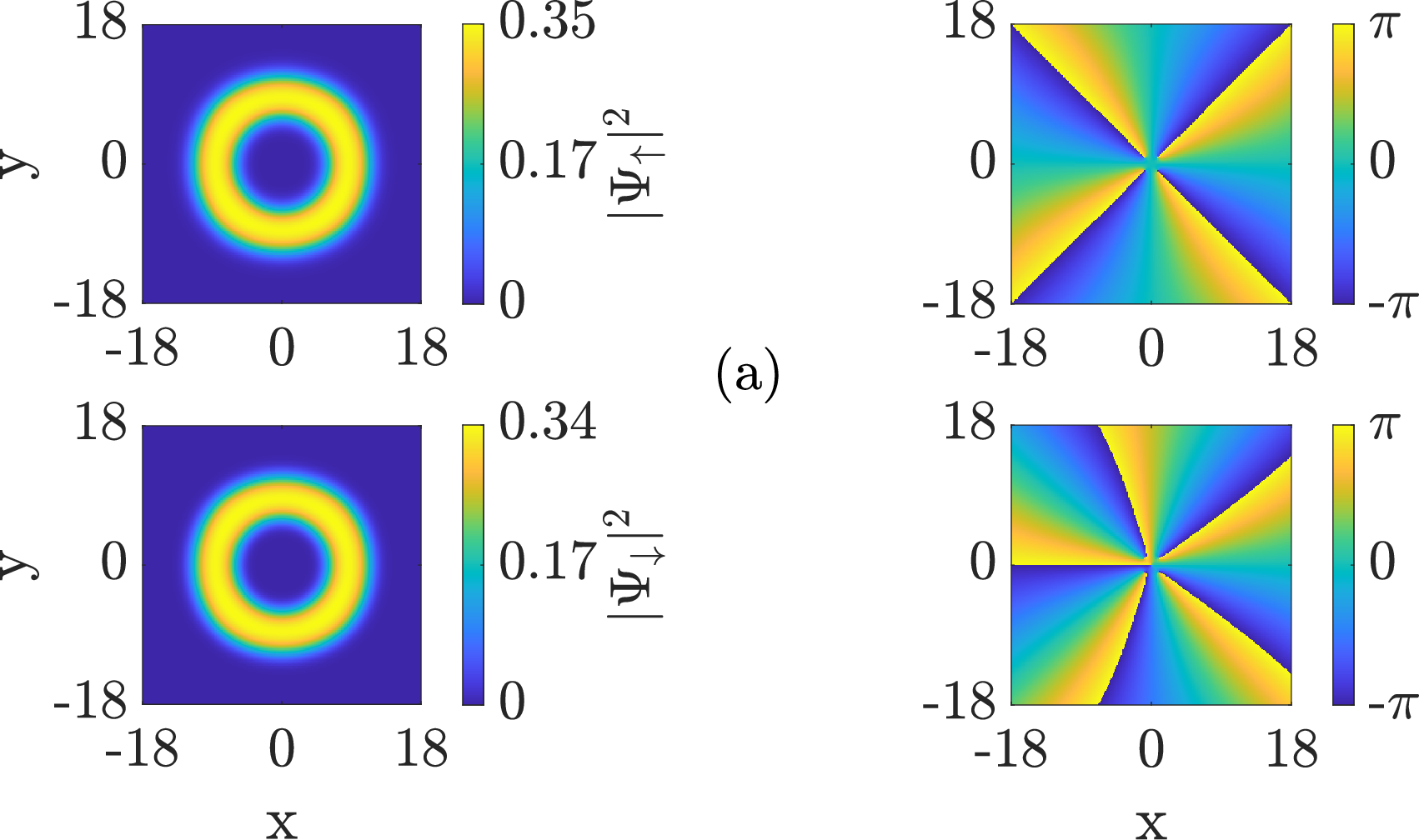}\\
\vspace{0.3\baselineskip}
\includegraphics[width=\columnwidth]{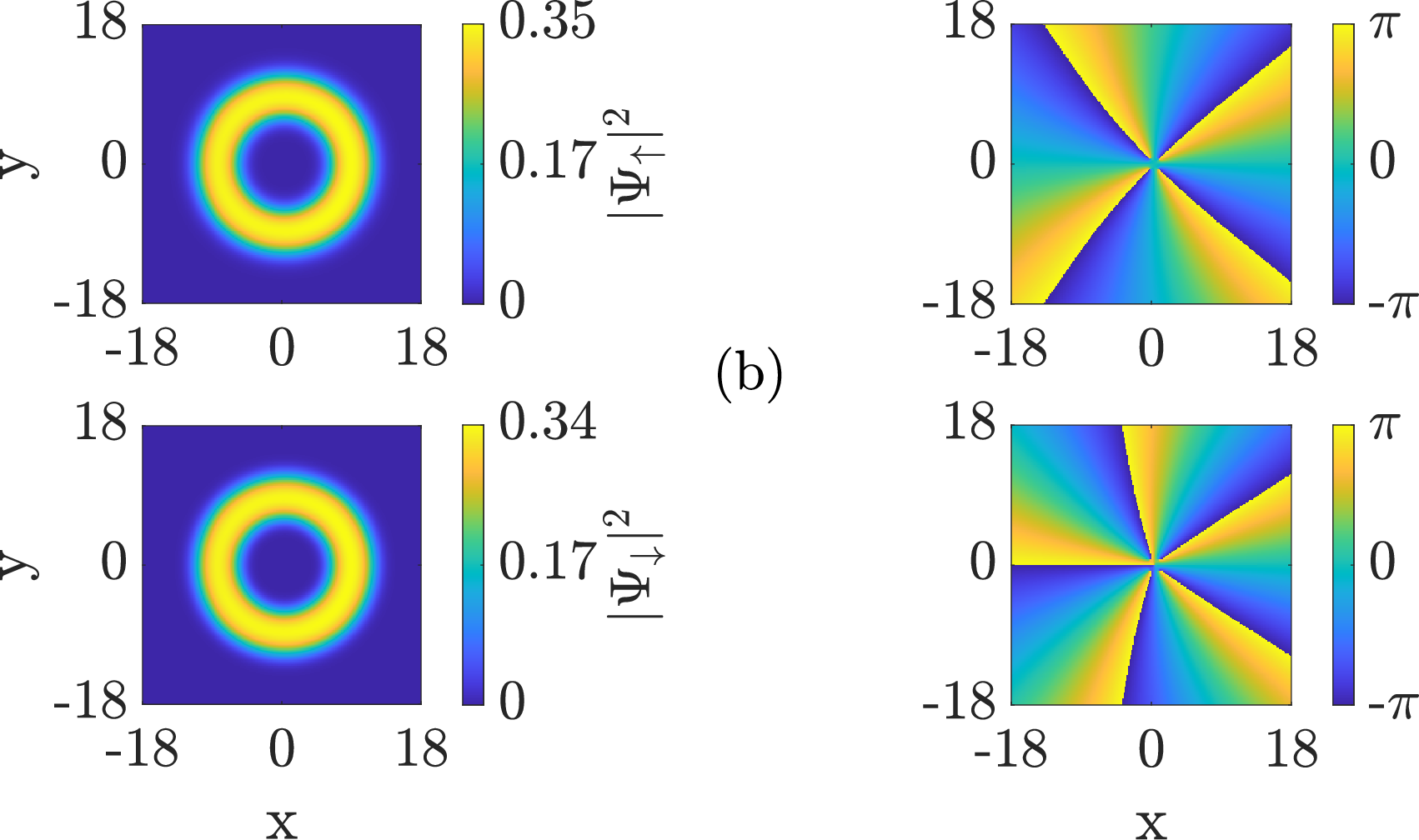}\\
\vspace{0.3\baselineskip}
\includegraphics[width=\columnwidth]{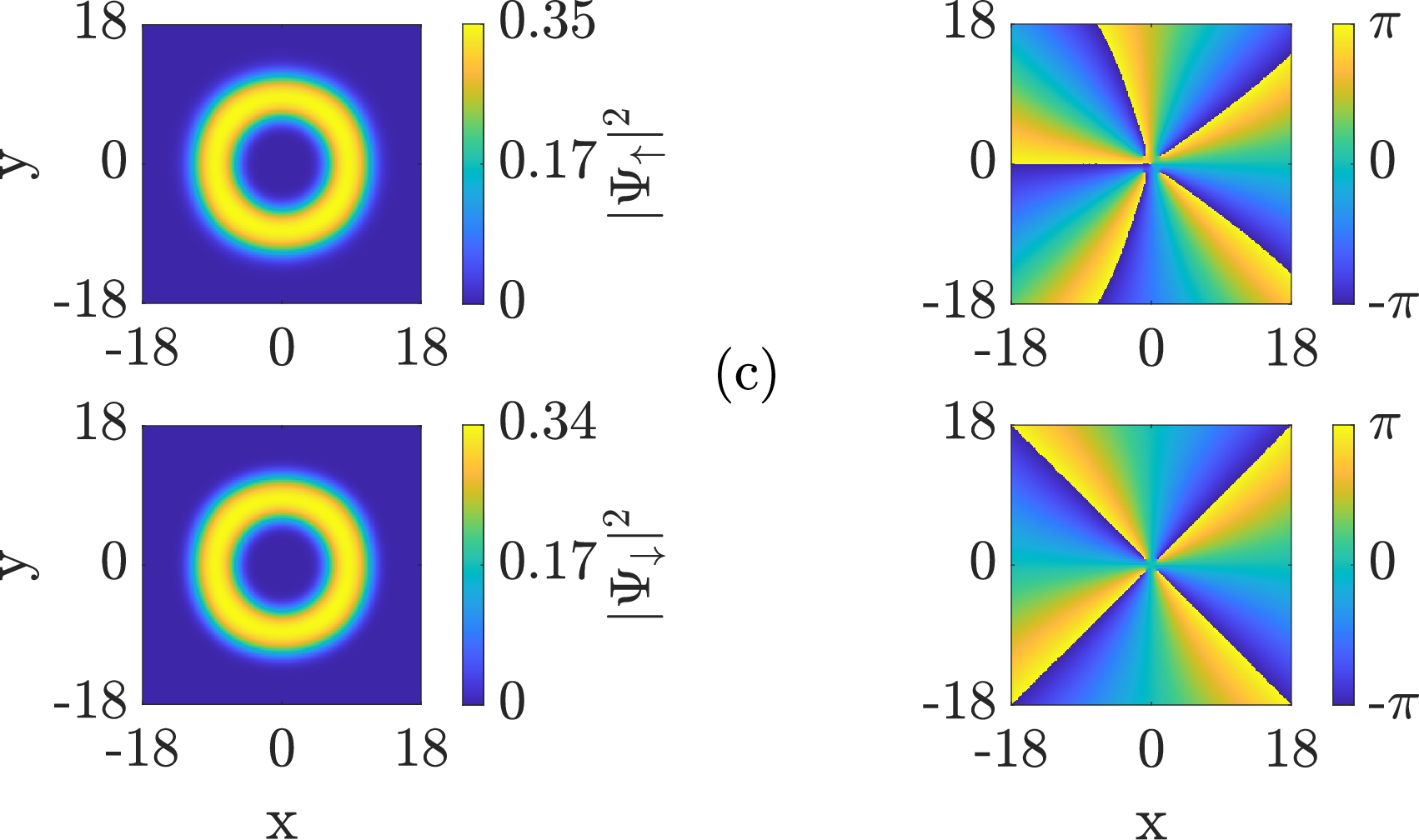}
\caption{(a)\textendash(c) The densities (left column, in units of $\Psi_0^2$) and the phases (right column) of the droplet order parameters, in the yrast state, for $N = 200$, $\omega = 0.05$, $\lambda = 0.05$, $\delta N = 5$, $D=25$, and (a) $\ell = 4.4875$, (b) $\ell = 4.5$, and (c) $\ell = 4.5125$. The unit of length is $x_0$.}
\addtocounter{figure}{-1}
\end{figure}

\begin{figure}[!ht]
\centering
\includegraphics[width=\columnwidth]{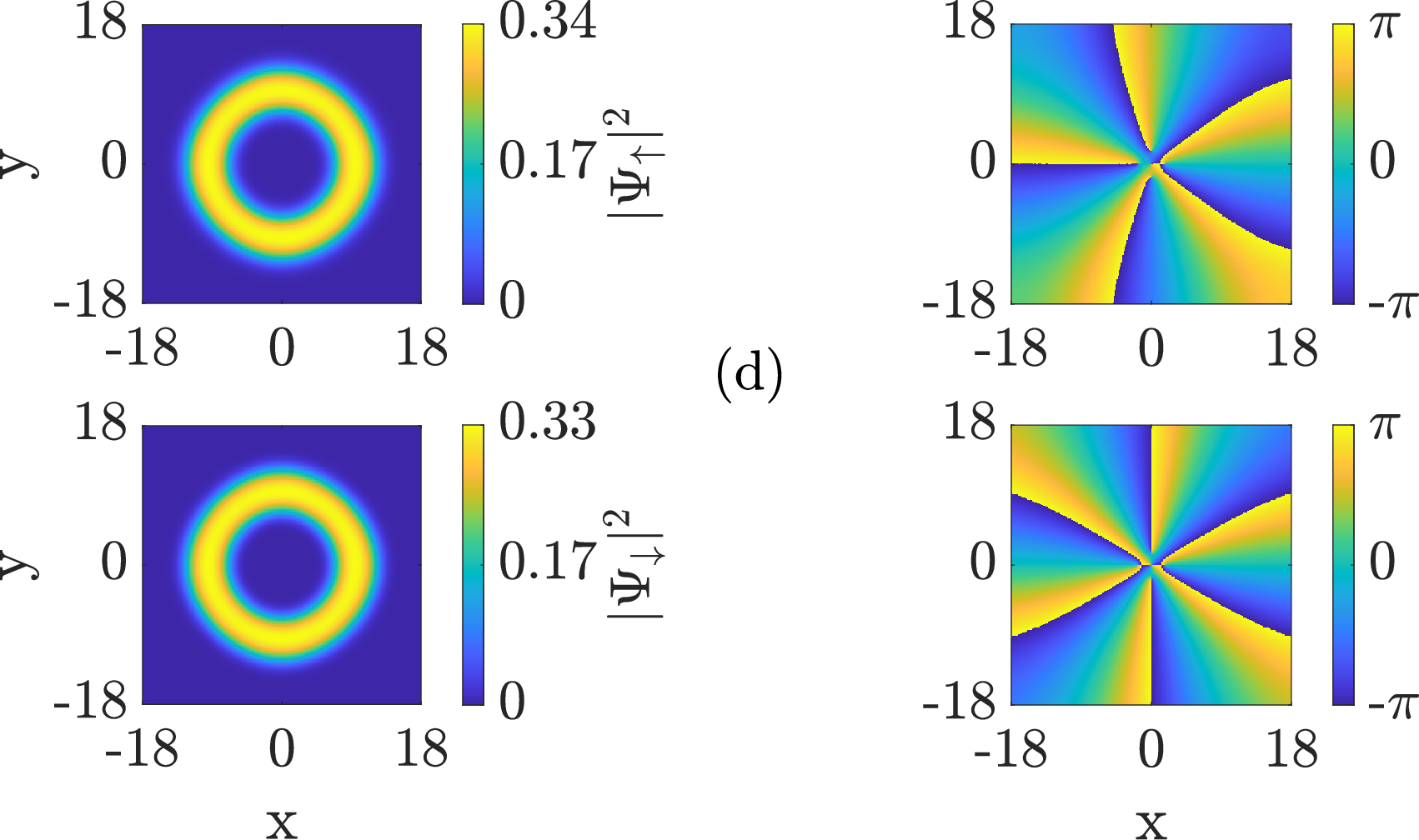}\\
\vspace{0.3\baselineskip}
\includegraphics[width=\columnwidth]{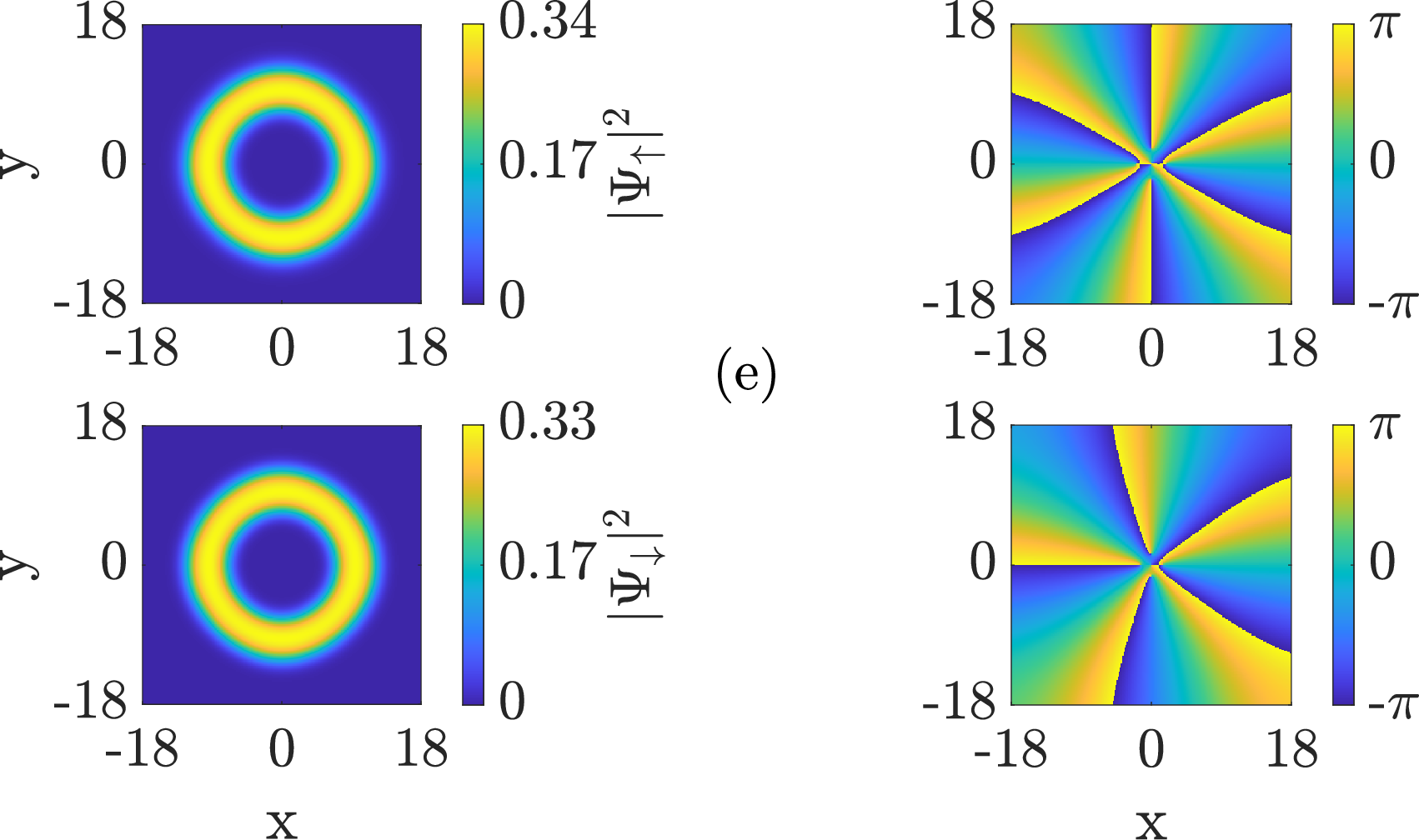}\\
\vspace{0.3\baselineskip}
\includegraphics[width=\columnwidth]{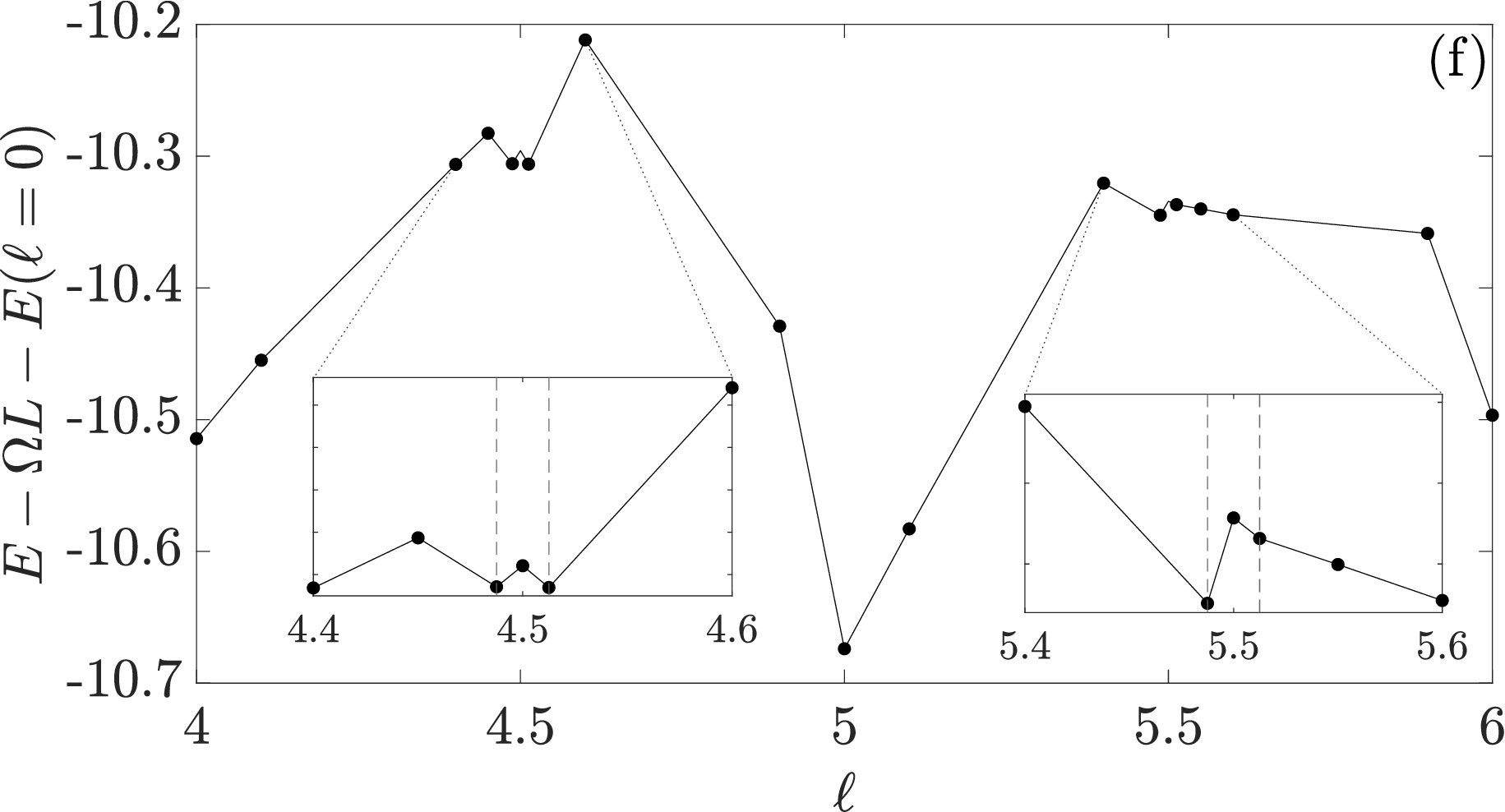}
\caption{(Cont.) (d), (e) Same as (a)\textendash(c) but for $\ell = 5.4875$ and $5.5125$, respectively. (f) The corresponding dispersion relation in the rotating frame, with $\Omega = 0.063$. The dashed vertical lines mark the values $L = 4N_\uparrow + 5N_\downarrow$, $5N_\uparrow + 4N_\downarrow$, $5N_\uparrow + 6N_\downarrow$, and $6N_\uparrow + 5N_\downarrow$. The unit of energy is $E_0$ and the unit of angular momentum is $\hbar$.}
\end{figure}

\subsection{Heterosymmetric multiply quantized vortex states of imbalanced droplets}

Finally, let us turn to the case of a rotating imbalanced droplet, with a main focus on how the heterosymmetric multiply quantized states are affected by a small population asymmetry. A qualitative difference in that case is that, in seeking such states, one is particularly interested in angular momentum values that constitute linear combinations of integer multiples of the two component populations, rather than half-integer values of $\ell$, as in the balanced case. In essence, for $L = m N_\uparrow + n N_\downarrow$, with $m$ and $n$ positive integers, a possible rotating state is an axially symmetric, heterosymmetric vortex state, with $m$ quanta of circulation in component $\uparrow$ and $n$ quanta in component $\downarrow$. As in the balanced case, we have found that when such configurations appear as yrast states they differ by one unit of circulation, that is, $n = m \pm 1$, and the corresponding angular momenta of such states are $L = m N_\uparrow + (m \pm 1) N_\downarrow$, neighboring a half-integer $\ell$ value.

To account for a small population imbalance in our example, we have set $\delta N = 5$ and have again varied the angular momentum per particle up to $\ell = 7$. Initially, we have found a heterosymmetric state  with a vortex only in component $\downarrow$ for $\ell = 0.4875$ ($L = N_\downarrow$), and a state with a vortex only in component $\uparrow$ for $\ell = 0.5125$ ($L = N_\uparrow$). This amounts to the effective lifting of the degeneracy of the $\ell = 0.5$ state that appears in the balanced system, as was also demonstrated for the case of harmonic confinement. For higher values of $\ell$, up to $\ell = 4.4$, we have found the yrast states to be phase-locked states throughout, qualitatively similar to those calculated for the balanced case, but with a small density difference among the two components, due to the population difference. Then, for $\ell = 4.4875$, that is, $L = 4N_\uparrow + 5N_\downarrow$, we have found a heterosymmetric vortex state with winding numbers $4$ in component $\uparrow$ and $5$ in component $\downarrow$, as presented in Fig.\,10(a). For $\ell = 4.5$ we have found a similar yrast state, which however slightly deviates from axial symmetry, as shown in Fig.\,10(b). This asymmetry is more apparent in the phase plot, where the discontinuity lines appear slightly shifted. This illustrates the fact that the heterosymmetric states we present here do not appear only for singular values of the angular momentum, but rather for a small range of angular momentum values, surrounding each axially symmetric state with $L = m N_\uparrow + (m \pm 1) N_\downarrow$, as the heterosymmetric states of Figs.\,3 and 7. For $\ell = 4.5125$, that is, $L = 5N_\uparrow + 4N_\downarrow$, we have found another heterosymmetric vortex state, now with winding numbers $5$ in component $\uparrow$ and $4$ in component $\downarrow$ [Fig.\,10(c)]. Again, we observe a lifting of the degeneracy between the states of Figs.\,10(a) and 10(c), as was mentioned in our treatment of the $\ell = 4.5$ state in the balanced droplet.

For $4.6 \leq \ell \leq 5.4$, we have again found the yrast states to be phase-locked, as in the balanced system. Interestingly, for $\ell = 5.4875$ ($L = 5N_\uparrow + 6N_\downarrow$) we have found a heterosymmetric, multiply quantized yrast state, with winding numbers $5$ in component $\uparrow$ and $6$ in component $\downarrow$ [Fig.\,10(d)], even though such a state does not appear as an yrast state in the balanced case. Similarly, for $\ell = 5.5125$ ($L = 6N_\uparrow + 5N_\downarrow$) the yrast state is again heterosymmetric, with winding numbers $6$ in component $\uparrow$ and $5$ in component $\downarrow$ [Fig.\,10(e)]. Therefore, we observe that the presence of a population imbalance has the effect of increasing the number of heterosymmetric states that appear in the yrast curve. It is also interesting to note that for $\ell = 5.5$, the yrast state is not heterosymmetric but rather phase-locked, similar to the state of Fig.\,9(d), but with the noted density difference between the components. Finally, for $5.6 \leq \ell \leq 7$, we have found that the yrast states are again phase-locked, similarly to the balanced case.

In Fig.\,10(f) we present the corresponding dispersion relation in the rotating frame, again for an angular velocity $\Omega = 0.063$ and on the range $4 \leq \ell \leq 6$. The insets focus in the two regions relevant to the heterosymmetric states. Overall, the result is similar to the balanced curve of Fig.\,9(f), although there is an energy increase compared to the balanced curve (not shown in this plot), which is owed to the contact energy term. An important difference is that we now observe two separate kinks/slope discontinuities for $\ell = 4.4875$ and $4.5125$ (instead of one for $\ell = 4.5$) which correspond to the heterosymmetric states of Figs.\,10(a) and 10(c). Additionally, we observe two kinks for $\ell = 5.4875$ and $5.5125$, corresponding to the heterosymmetric states of Figs.\,10(d) and 10(e), whereas the balanced curve has a continuous slope in this region of $\ell$ values. The second kink in particular, for $\ell = 5.5125$, is only slightly apparent in the plot, due to its energy being very close that of the corresponding phase-locked state. We note here that the transitions between different heterosymmetric states, as well as between heterosymmetric and phase-locked states, are discontinuous. This means that between $\ell = 5.4$ and $5.6$ there exist four discontinuous transitions, all in a relatively small range of angular momentum values. Consequently, in this range of $\ell$ values there are six discontinuities in the slope of the curve: four corresponding to the phase transitions (which are not displayed in the plot), and two corresponding to the axially symmetric vortex states. Conversely, the curve itself is continuous.

\section{Discussion and summary}

In the present paper we investigated the rotational response of two-dimensional, confined quantum droplets, which arise in a mixture of two distinguishable Bose-Einstein condensed gases. We have considered both cases of harmonic and anharmonic confinement. In addition, we treated droplets with equal populations in the two components, as well as droplets with a small population imbalance. We have gone beyond the usual model, which reduces the system to the calculation of a single order parameter that is shared between the two components, instead treating the order parameter of each component separately, by solving the pair of coupled differential equations that describe the system.

We have focused on the yrast states, that is, the lowest-energy states for each value of the angular momentum, and we have placed a major emphasis on states where the two components are excited in a topologically different manner. More specifically, we have uncovered a number of heterosymmetric vortex yrast states, where the two components carry different vorticities. Importantly, such states are missed by the single-order-parameter model.

In the case of harmonic confinement, the heterosymmetric states we have found take the form of a vortex in only one component, accompanied by a density decrease/partially filled core in the other component. The requirement for such states to appear as the lowest-energy state is that the droplet is under sufficiently strong confinement. In the opposite case, the droplet exists in a phase-locked state, where the two components are excited in the same manner. Interestingly, we have found that while the kinetic, potential and contact energy terms increase in the heterosymmetric state (compared to the phase-locked one), the logarithmic energy term decreases. This competition determines the crossover to the heterosymmetric state, as a function of the trapping frequency, as we have demonstrated with a semianalytic calculation. Since the logarithmic energy term captures the beyond-mean-field effects that are prominent in the droplet system, the appearance of such heterosymmetric states in the yrast curve can therefore be regarded as a manifestation of beyond-mean-field physics. It is also interesting to note that such states appear as yrast states even in the case of zero mass-, interaction- and population-asymmetry (which in our specific system corresponds to the balanced case). Thus, the appearance of heterosymmetric yrast states cannot be attributed solely to some manifest asymmetry in the system.

In the case of anharmonic confinement, heterosymmetric vortex yrast states are also present, and indeed more numerous. This is owed to the propensity of the anharmonic potential to support multiply quantized vortex states, for sufficiently high values of the angular momentum/angular velocity. Apart from states with the same form as the heterosymmetric states we have found in the case of harmonic confinement, here there also exist heterosymmetric multiply quantized vortex states, with the circulation differing by one unit between the two components.

Contrary to the single-order-parameter model, the use of the present model also allows for the treatment of nonzero imbalance in the quantum droplet system. In this work, we have investigated the effects of a small population imbalance, with regards to both the phase-locked and the heterosymmetric states. In both cases, we have noted a small increase in the energy of the droplet, owed to the positive contribution of the contact energy term (which is quadratic in the densities difference). In the case of phase-locked states, the differences to the balanced case are simply quantitative, i.e., the aforementioned energy increase, as well as a corresponding density difference between the two components.

In the case of heterosymmetric states, the effects of a population imbalance are more profound. Apart from the energy difference, the presence of a population asymmetry induces a qualitative change in the structure of the yrast curve. In a perfectly balanced droplet, for each heterosymmetric state there exists a second, essentially identical state, but with the vorticities interchanged between the two components. These states are degenerate, which is a manifestation of the spontaneous breaking of the $\mathbb{Z}_2$ symmetry underlying the balanced droplet. Under the presence of a nonzero imbalance, which explicitly breaks the $\mathbb{Z}_2$ symmetry, such two states correspond to different values of the angular momentum and have different energies. Thus, we observe that the presence of a population imbalance lifts the degeneracy of the heterosymmetric states. In addition, we have found that the introduction of imbalance has the effect of making the presence of heterosymmetric states in the yrast curve more ubiquitous, at least for anharmonic confinement. This is demonstrated by the appearance of a pair of heterosymmetric yrast states in the imbalanced, anharmonically confined droplet, while they are absent in the corresponding balanced yrast curve.

Clearly, part of the value of the present study lies in determining the range of accuracy of the simpler, single-order-parameter model, which is very often employed. Naturally, if one wishes to study an imbalanced droplet, only the present full model is applicable. Even in the balanced case, significant care should be taken when investigating angular momentum values that are candidates of supporting heterosymmetric yrast states, especially in cases of strong confinement, as the single-order-parameter model does not reproduce such configurations. For a harmonic potential, we have demonstrated this for a range of values of the angular momentum per particle close to $0.5$, where we have identified a series of heterosymmetric states. For an anharmonic potential, we have found that heterosymmetric yrast states additionally appear in certain ranges of the angular momentum per particle, close to half-integer values. Therefore, when investigating the lowest-energy configuration for such values of the angular momentum, the full model will be invaluable in determining the correct state, when this is a heterosymmetric state.

The results of this work have been presented in the dimensionless units defined in Sec.\,II. Let us now make a connection with the physical units, which are more relevant to actual experiments. To begin with, $D = 25$ corresponds to $a^{\rm 3D} \sim 10.1$ nm, $a_{\uparrow\downarrow}^{\rm 3D} \sim -10.0$ nm, and $a_z \sim 0.1$ $\mu$m, as per Eq.\,(\ref{logarithm}). We stress that the dependence on the intra- and intercomponent 3D scattering lengths, and on the oscillator length in the transverse confinement direction, make $D$ an experimentally tunable parameter. Then, according to Eq.\,(\ref{atom_number_unit}), we have $N_0 \approx 50$. For $N = 200$, this corresponds to approximately $10000$ atoms in an experiment. Regarding the population imbalance, $\delta N = 5$ corresponds to approximately $250$ more atoms in the majority component. In addition, according to Eq.(\ref{length_unit}), the unit of length $x_0$ is on the order of $1$ $\mu$m, and the size of the droplets we have investigated is on the order of tens of $\mu$m. Finally, a typical value of the trapping frequency is on the order of hundreds of Hz, while a typical value of the (dimensionless) anharmonicity parameter is $\sim 10^{-2}$ \cite{Dalibard}.

The calculations in this work have been carried out under the constraint of fixed angular momentum. However, the knowledge of the dispersion relation, that is, the yrast energy as a function of the angular momentum, allows us to determine how the system behaves for a fixed value of the angular velocity, instead. This is achieved by producing the dispersion relation in the rotating frame, for a given value of the angular velocity, and determining the global minimum of the curve. Interestingly, for the parameter values we have investigated in this work, the heterosymmetric yrast states we have uncovered do not appear as global minima for any value of the angular velocity in the rotating frame. This means that our results regarding fixed values of the angular velocity in Refs.\,\cite{NKO1,NKO2} still hold in the light of these new results. Conversely, working with fixed values of the angular momentum instead, we have seen that the results of these studies are modified by the presence of the heterosymmetric states. We stress that, experimentally, one may choose to work with a fixed value of the angular momentum, rather than the angular velocity.

The above observation may appear to indicate that the single-order-parameter model is a reliable tool when working with a fixed angular velocity. However, we have to stress that this reported absence of heterosymmetric configurations in the lowest-energy state\textemdash for fixed values of the angular velocity\textemdash only pertains to the specific parameter values we have considered in this work. A more extensive investigation of this problem could be the subject of a future work.

Finally, let us comment on some additional possible extensions of the present work. A natural avenue would be to investigate the presence of heterosymmetric states in two-dimensional, heteronuclear and/or interaction-asymmetric quantum droplets. Such states have attracted attention in the literature for the three-dimensional case \cite{Caldara,Poparic}. Although the asymmetry of masses or intracomponent interaction strengths is not a prerequisite for heterosymmetric states to arise in the yrast curve, as the present work has demonstrated, it is nevertheless interesting to study how the additional asymmetry may modify or even enrich such exotic structures in this intriguing, quantum phase of matter.

\section*{Acknowledgment}

S. N. acknowledges support from the Hellenic Foundation for Research and Innovation (HFRI) under the 5th Call for HFRI PhD Fellowships.

\end{document}